\documentclass[twocolumn,showpacs,preprintnumbers,amsmath,amssymb]{revtex4}

\usepackage{graphicx}
\usepackage{dcolumn}
\usepackage{bm}
\usepackage{amsmath}

\begin{document}

\title{Flow equation approach to heavy fermion systems}

\author{Karsten Meyer}
\affiliation{%
Institut f\"ur Theoretische Physik, Technische Universit\"at Dresden, 
01062 Dresden, Germany
}%

\date{\today}

\begin{abstract}
We use Wegner's flow equation method to investigate the infinite-$U$ periodic
Anderson model. We show that this method poses a new approach to the
description of heavy fermion behaviour. Within this scheme we derive an
effective Hamiltonian in which the $c$ and $f$ 
degrees of freedom are decoupled. By evaluating one-particle energies as well
as correlation functions we find an electronic structure which comprises two
gapped quasiparticle bands. We also address the lattice Kondo
temperature, which shows a typical exponential dependence on the hybridisation
energy. This energy scale exhibits a significant decrease compared to that
of the single impurity Anderson model.  
\end{abstract}
\maketitle
\section{Introduction}
Heavy fermion systems (HFS) have triggered a growing interest in both
experimental and theoretical physics over the last decades. They are
mainly based upon lanthanide or actinide compounds, thus their physics
is dominated by the interplay of itinerant ($3d$) electrons and
rather localised electrons from partly filled $4f$ or $5f$ electron shells.
These systems generally posses rich phase diagrams, and 
they show intriguing physical phenomena, such as superconductivity,
long-range magnetic order, Fermi-liquid and non-Fermi-liquid
behaviour, and in many cases magnetism and superconductivity
coexist. Despite much theoretical effort, many open questions remain even for
the paramagnetic metallic phase. At low temperatures this
phase is generally 
characterised by the existence of heavy quasiparticles, which lead to a
strongly enhanced specific heat coefficient as well as an enhanced
susceptibility. Experiments suggest that the effective mass of the
quasiparticles is in the region of several hundreds up to thousand
bare electron masses.
On a microscopic basis, HFS can be described by the periodic Anderson
model (PAM) and the Kondo lattice model (KLM). Whereas the first
describes a system of localised $f$ electrons and conduction
electrons, which interact via a hybridisation, the latter yields
the microscopic picture of a periodic range of magnetic moments in a
metallic host. In the so-called Kondo limit the KLM can be regarded as
an effective model for the PAM. The physics of both models is
associated with the Kondo effect \cite{Hewson1993}, which was
initially examined in connection with a single magnetic impurity
interacting with a sea of free conduction electrons. The typical
energy scale for this interaction is provided by the Kondo temperature
$T_\mathrm{K}$, which shows an exponential dependence on the
hybridisation strength. This relation cannot be understood in terms of 
conventional perturbation theory. Beside systems that involve $f$ electrons, 
heavy fermion behaviour has also been observed in
$\mathrm{LiV_2O_4}$~\cite{Kondo1997}. It is believed that the occurrence of a
heavy Fermi liquid state is driven by the frustration of the intersite spin
couplings~\cite{Hopkinson2002}.\par 
About 10 years ago Wegner \cite{Wegner1994} and independently G\l azek
and Wilson \cite{Glazek1993} proposed a scheme with the objective to
diagonalise or block-diagonalise Hamiltonians on the basis of a
continuous unitary transformation. This method is known as the flow
equation method and has been used for the theoretical description of
a great variety of physical systems.
In a recent application this method was used to investigate the Hubbard model, a 
paradigm for a  strongly correlated lattice system
\cite{Grote2002}. Although the authors treated the system  on the basis
of perturbation theory, they obtained excellent results in the
regions of moderate and even strong coupling. In this paper we apply the flow
equation method to the infinite-$U$ PAM in order to describe the physics of
the paramagnetic metallic phase of HFS. Thereby we
do not consider perturbative arguments with respect to the hybridisation
strength. We focus on the
electronic structure, which suggests the existence of heavy
quasiparticles, and show that the lattice Kondo temperature features an
exponential dependence on the hybridisation strength. This paper starts off
with some general remarks regarding the method and the microscopic
model. Afterwards we  discuss the application of the method as an approach to
the PAM. The last part is devoted to a detailed discussion of our findings. 
\section{Model and Method}
\subsection{Periodic Anderson Model}
The PAM is the standard model for the description of heavy fermion
systems. Supposing the Coulomb repulsion of the local 
$f$ electrons being the dominant energy scale in the system, this model can
be written in the form
\begin{align}
\label{PAM}
H_\mathrm{PAM}=&
\sum_{\mathbf{k}\sigma}\varepsilon_\mathbf{k}\,
c_{\mathbf{k}\sigma}^\dagger c_{\mathbf{k}\sigma}^{\phantom{\dagger}}
+ \sum_{i\sigma}\varepsilon_f
\,\hat{f}_{i\sigma}^\dagger
\hat{f}_{i\sigma}^{\phantom{\dagger}}\nonumber\\
+&\frac{1}{\sqrt{N}}\sum_{\mathbf{k}i\sigma}V_\mathbf{k}\,
\bigl(c_{\mathbf{k}\sigma}^\dagger
\hat{f}_{i\sigma}^{\phantom{\dagger}}\,e^{-i\mathbf{kR}_i}+
\hat{f}_{i\sigma}^\dagger
c_{\mathbf{k}\sigma}^{\phantom{\dagger}}\,e^{i\mathbf{kR}_i}\bigr)
\end{align}
In this representation $\hat{f}_{i\sigma}^\dagger$ creates an $f$ electron on
site $i$ with spin $\sigma$. Due to the infinitely strong Coulomb repulsion of
two $f$ electrons doubly occupied sites are 
explicitly excluded. This is done by the use of Hubbard operators rather than
usual fermionic creation and destruction operators:
\begin{align}
\hat{f}_{i\sigma}^\dagger=f_{i\sigma}^\dagger\,\prod_{\sigma^\prime} \bigl(
1-n_{i\sigma^\prime}^f\bigr)=f_{i\sigma}^\dagger P_i^0\\
\hat{f}_{i\sigma}^{\phantom{\dagger}}=\prod_{\sigma^\prime} \bigl(
1-n_{i\sigma^\prime}^f\bigr)\,f_{i\sigma}^{\phantom{\dagger}}= 
P_i^0 f_{i\sigma}^{\phantom{\dagger}}
\end{align}
Here $n_{i\sigma}^f$ denotes the number operator for the $f$
electrons. The essential property of the Hubbard operators is that the
creation operator $\hat{f}_{i\sigma}^\dagger$ only acts on an empty site $i$, and
the  operator $\hat{f}_{i\sigma}^{\phantom{\dagger}}$ only annihilates a singly
occupied state. The operator $P_i^0$ stands for the projector on the empty
$f$ site $i$. In this context the physical subspace merely comprises singly
occupied and empty $f$-sites. As a consequence the completeness relation 
$P_i^0+\sum_\sigma
\hat{f}_{i\sigma}^\dagger\hat{f}_{i\sigma}^{\phantom{\dagger}}=\mathbf{1} $
holds, and the 
operators obey the following anti-commutation relations
\begin{equation}\label{eq:com_rel}
  \bigl\{
  \hat{f}_{i\sigma}^\dagger,\hat{f}_{i'\sigma'}^{\phantom{\dagger}}\bigr\}_+=
  \delta_{ii'}\bigl( \delta_{\sigma\sigma'}\,
  \hat{f}_{i\sigma}^{\phantom{\dagger}} \hat{f}_{i\sigma}^\dagger +
  \hat{f}_{i\sigma}^\dagger\hat{f}_{i\sigma'}^{\phantom{\dagger}}\bigr)
\end{equation}
The Fourier transform of the Hubbard operators can be defined in the usual way
as $\hat{f}_{\mathbf{k}\sigma}^\dagger=1/\sqrt{N}\sum_i
\exp(i\mathbf{kR}_i)\hat{f}_{i\sigma}^\dagger$, where $N$ is the number of
lattice sites. However, the above introduced commutation relations for the
Hubbard operators in the local picture result in rather complex commutation
relations for the operators $\hat{f}_{\mathbf{k}\sigma}^\dagger$ and
$\hat{f}_{\mathbf{k}\sigma}^{\phantom{\dagger}}$. 
In Eq.~\eqref{PAM} 
$c_{\mathbf{k}\sigma}^\dagger$ creates a conduction 
electron with momentum $\mathbf{k}$ and spin $\sigma$. Whereas the $f$
electrons are localised, the conduction electrons possess a dispersion
$\varepsilon_\mathbf{k}$. The two subsystems are connected via a
hybridisation of the strength $V_\mathbf{k}$. Although, in general, a
dispersive hybridisation may be examined, we here focus on the case
$V_\mathbf{k}=\mathrm{const}$, i.~e. an interaction which is restricted to
electrons on the same lattice site only. As a further simplification we only
consider a hybridisation between $c$ and $f$ electrons within the same spin
channel. \par
The model \eqref{PAM} has been in the focus of a great variety of theoretical
investigations. One of the most descriptive ones is the slave-boson mean-field (SBMF)
theory \cite{SlaveBoson}. Here the Hubbard operators are treated by the
introduction of auxiliary boson fields, which ensure the conservation of the
commutation relations. On the other hand, the configuration space of the
particles must be restricted due to their bosonic character. This can be
accomplished by considering the constraint with the help of additional
couplings in the Hamiltonian. The used Lagrange multipliers have to be determined
self-consistently. For this model the problem can be solved in the limit
$\nu_f\rightarrow\infty$, where the latter quantity denotes the spin
degeneracy of the $f$ orbitals. A quite similar treatment of the model is
provided by the Gutzwiller variational method \cite{Rice1985}, which can also be
formulated in terms of auxiliary bosons \cite{Kotliar1986}. The intriguing result
of this approach is an instability of the ground state with respect to
ferromagnetic ordering for the case $\nu_f=2$, and the Kondo temperature is
significantly enhanced compared to the single impurity case.\par
Though the analytical approaches to the PAM have given insight to various
aspects of heavy fermion systems, they generally have to deal with more or
less rude simplifications. In the last decade a range of numerical solutions
emerged. A great number of recent
works are based on the Dynamical mean-field theory
(DMFT)~\cite{DMFT}. Here the PAM can be 
mapped onto an effective single impurity Anderson model (SIAM). This
approach proves to be exact in infinite dimensions, and it is considered to
deliver good results even for three-dimensional systems. The major task of
this scheme is a self-consistent numerical solution of the impurity problem,
whereby a number of methods, such as Wilson's numerical renormalisation group
(NRG)~\cite{Pruschke2000}, Quantum Monte Carlo (QMC)
simulations~\cite{Tahvildar1998} and iterated perturbations theory
(ITP)~\cite{Vidhyadhiraja2000}, have been used. Other numerical  
approaches comprise the diagonalisation of small clusters~\cite{Tsutsui1996}
and QMC~\cite{Groeber1999} for the full system.
As the numerical results of Sec.~\ref{sec:results} are
obtained for a one-dimensional system, we should mention that this
system and the corresponding KLM have been extensively studied in the
framework of the Density Matrix Renormalisation Group
(DMRG)~\cite{Tsunetsugu1997}. Various studies of the
KLM suggest this system being in the universality class of Luttinger
liquids. The
phase diagram of the one-dimensional PAM was subject of a series of
recent works~\cite{Guerrero}.
\par
Some basic properties of the model Eq.~\eqref{PAM} have also been discussed
in the framework of a renormalisation approach by Becker \textit{et
  al.}~\cite{Becker2002}, which incorporates the use of 
hard cut-off functions. However, the authors use
perturbation theory with respect to the hybridisation strength and neglect
processes that connect different $f$ sites, 
which is not sufficient to obtain the proper heavy fermion behaviour.
\subsection{Flow equation method}\label{sec:FE}
The flow equation method is based on
the formulation of a continuous unitary transformation.
For this reason in various works this approach is refered to as CUT.
The transformation is represented by
a unitary Operator $U(\ell)$, for a given Hamiltonian $H$. Here $\ell$ denotes
the flow parameter~\cite{Flowparameter} and the according transformed
Hamiltonian is simply given by 
$H(\ell)=U(\ell)HU^\dagger(\ell)$. By definition the operator
$U(1)$ stands for the identity operator and the fully transformed
Hamiltonian $H(\ell=\infty)$ is written as $H^\ast$. The latter
may be regarded as an effective Hamiltonian, and due to the unitarity of
the mapping it possesses the same system of eigenenergies as $H$. The flow of
$H(\ell)$ can be conveniently formulated in its differential
representation
\begin{equation}\label{eq:flow_Hamiltonian}
  \frac{\mathrm{d}H(\ell)}{\mathrm{d}\ell}=\bigl[\eta(\ell),H(\ell)\bigr]
\end{equation}
and it is governed by the generator 
\begin{equation}
  \eta(\ell)=\frac{\mathrm{d}U(\ell)}{\mathrm{d}\ell} U^\dagger(\ell)=-\eta^\dagger(\ell)
\end{equation}
This operator can be chosen to map $H$ onto a much simpler
Hamiltonian. If one starts out, for instance, with a Hamiltonian of
the structure $H=H_0+H_1$, where $H_0$ represents a system of free particles
and $H_1$ an interaction, the choice
\begin{equation}\label{eq:wegner}
 \eta=\bigl[H_0,H_1\bigr]
\end{equation}
indeed triggers the interaction part to vanish in the limit
$\ell\rightarrow\infty$. This can be proved by some simple calculation
and Eq.~\eqref{eq:wegner} represents the original choice of the generator
by Wegner~\cite{Wegner1994}. In this context the resulting effective Hamiltonian is
diagonal or block-diagonal. Beside Wegner's generator, in the recent
years a variety of diverse generators has been used. As we have mentioned in
the introductory section, an independent approach formulated by G\l azek
and Wilson was made, in which the Hamiltonian is diagonalised by
integrating out higher energy contributions. Another choice for the generator
was proposed by Knetter and Uhrig~\cite{Knetter2000}. It has the intriguing
feature that it allows the construction of a particle-number conserving
effective Hamiltonian.\par 
Independent of the choice of any particular generator, the
transformation can as well be applied to formulate the flow of observables. Similar
to the structure we have derived for the Hamiltonian,
an arbitrary operator is transformed as
\begin{equation}
  \frac{\mathrm{d}A(\ell)}{\mathrm{d}\ell}=\bigl[\eta(\ell),A(\ell)\bigr]
\end{equation} 
As a result of this we can take advantage of the invariance of
expectation values and correlation functions under unitary
transformations. In this sense these quantities can be evaluated with
respect to the much simpler Hamiltonian $H^\ast$ by the simultaneous
transformation of the operators. If we regard for example the
expectation value of an observable $A$, the relation
\begin{equation}
  \langle A \rangle = \frac{\mathrm{Tr} A e^{-\beta H}}{Z}=
  \frac{\mathrm{Tr} A^\ast e^{-\beta H^\ast}}{Z}
\end{equation}
holds, where $Z$ is the partition function.\par
The above considerations are instructive for the evaluation
of Green's functions. If we address, for instance, the retarded
Green's function $\langle\langle A(t);B(t')\rangle\rangle=-i
\theta(t-t')\langle\{A(t),B(t')\}_+\rangle$, we can use the fact that these
quantities are invariant with regard to a unitary transformation. This
allows us to evaluate the expectation value and the Heisenberg time
dependence of the operators with respect to the transformed
Hamiltonian $H^\ast$, if we use the transformed
operators $A^\ast$ and $B^\ast$ at the same time. In terms of the Fourier transformed
Green's function this can be simply denoted as $\langle\langle
A;B\rangle\rangle(\omega + i\delta)=
{}^\ast\langle\langle A^\ast;B^\ast\rangle\rangle^\ast(\omega + i\delta)$.  
\section{Flow equations for the Periodic Anderson Model}
After having discussed the general aspects of the model and the method we now 
focus on the application of the flow equation approach to the infinite-$U$
PAM. This section is organised as follows. 
First we introduce an appropriate generator with the objective to map the model onto a 
much simpler one, and subsequently we discuss the flow of the Hamiltonian. In a second 
step the transformation of observables is examined and the evaluation of
expectation values is discussed. The last part of this section then is devoted
to the solution of the resulting flow equations. 
\subsection{Flow of the Hamiltonian}\label{sec:flow}
As we have mentioned in the introductory remarks regarding the flow equation method an
appropriate choice of the generator is provided by \eqref{eq:wegner}. The aim of our
treatment of the PAM is the decoupling of the distinct types of electrons, 
namely the $c$ electrons and the $f$ electrons. They are connected by a hybridisation which 
is characterised by the energy $V_{\mathbf{k}}$. If we identify this contribution with
the interacting part of the Hamiltonian, we recognise that the remaining contribution
is not diagonal in the usual sense. This is due to the representation of the
$f$ electrons by Hubbard operators and results from the incorporation of their
strong Coulomb interaction. 
Consequently, we use for the generator the more general form
\begin{equation}\label{eq:generator}
 \eta(\ell)=\sum_{\mathbf{k}\sigma} \eta_{\mathbf{k}}(\ell)\,
 \bigl(c_{\mathbf{k}\sigma}^\dagger
 \hat{f}_{\mathbf{k}\sigma}^{\phantom{\dagger}}-
 \hat{f}_{\mathbf{k}\sigma}^\dagger 
 c_{\mathbf{k}\sigma}^{\phantom{\dagger}})
\end{equation}
The coefficient $\eta_{\mathbf{k}}$ has to be determined later. A similar definition has
already been used for the flow equations analysis of related
models~\cite{Kehrein1996,Kehrein1996a}. \par 
A first remarkable result of the above choice for the generator is
the generation an $f$ electron hopping of the form
\begin{equation}
H_{ff}=\sum_{i\neq j\sigma}t_{ij}(\ell)\,\hat{f}_{i\sigma}^\dagger
\hat{f}_{j\sigma}^{\phantom{\dagger}}
\end{equation}
as a consequence of the evaluation of the commutator in Eq.~\eqref{eq:flow_Hamiltonian}.
A detailed discussion of all contributions is given below. The dispersion of the 
$f$ electrons is obtained by the Fourier transform of the hopping matrix element $t_{ij}$
with the exclusion of local processes
\begin{equation}
\Delta_\mathbf{k}(\ell)=\frac{1}{N}\sum_{i\neq j}t_{ij}(\ell)\,
e^{-i\mathbf{k}(\mathbf{R}_i-\mathbf{R}_j)}
\end{equation}
With this additional interaction term the Hamiltonian of the flow results in the following 
form:
\begin{align}\label{eq:PAM_flow}
H_\mathrm{PAM}(\ell)=&
\sum_{\mathbf{k}\sigma}\varepsilon_\mathbf{k}(\ell)\,
c_{\mathbf{k}\sigma}^\dagger c_{\mathbf{k}\sigma}^{\phantom{\dagger}}
+\sum_{\mathbf{k}\sigma}\bigl(\varepsilon_f(\ell) +
\Delta_\mathbf{k}(\ell)\bigr)\,\hat{f}_{\mathbf{k}\sigma}^\dagger
\hat{f}_{\mathbf{k}\sigma}^{\phantom{\dagger}}\nonumber\\
&+\sum_{\mathbf{k}\sigma}V_\mathbf{k}(\ell)\,
\bigl(c_{\mathbf{k}\sigma}^\dagger
\hat{f}_{\mathbf{k}\sigma}^{\phantom{\dagger}}+
\hat{f}_{\mathbf{k}\sigma}^\dagger
c_{\mathbf{k}\sigma}^{\phantom{\dagger}}\bigr)
+E(\ell)
\end{align}
Besides the hopping of the $f$ electrons an additional energy shift $E(\ell)$ was
introduced.\par
As a general convention for the upcoming considerations we suppress the
explicit declaration of the $\ell$-dependence of all energies. We only use
it in some cases in order to avoid confusions. Furthermore, we use for a
certain quantity $Q$ the representation $Q^\mathrm{i}$ for its initial value,
i.~e. $Q(\ell=0)$, and $Q^\ast$ for the limit $\ell\rightarrow\infty$.\par
Eq.~\ref{eq:flow_Hamiltonian} poses the central relation of the
flow equation method, thus our task is the evaluation of the commutator of the 
generator \eqref{eq:generator} and the Hamiltonian \eqref{eq:PAM_flow}. As
the scheme suggests the resulting 
contributions determine the flow of the Hamiltonian. The commutator
can be easily calculated by making use of the commutator relation
\eqref{eq:com_rel} for the Hubbard operators. By doing so we arrive at the
following result: 
\begin{align}\label{eq:commutator}
\bigl[\eta,H\bigr]&=
-\sum_{\mathbf{k}\sigma}\eta_\mathbf{k}\,\bigl(\varepsilon_\mathbf{k}
-\varepsilon_f\bigr)\,\bigl(
c_{\mathbf{k}\sigma}^\dagger
\hat{f}_{\mathbf{k}\sigma}^{\phantom{\dagger}}
+ \hat{f}_{\mathbf{k}\sigma}^\dagger
c_{\mathbf{k}\sigma}^{\phantom{\dagger}}
\bigr)
\nonumber\\
&+\frac{1}{\sqrt{N}}\sum_{\mathbf{k}\sigma}\sum_{i\neq j}
\eta_\mathbf{k}\,t_{ij}\,e^{-i\mathbf{kR}_i} 
P_i^0 
c_{\mathbf{k}\sigma}^\dagger
\hat{f}_{j\sigma}^{\phantom{\dagger}}
+\text{h.~c.}\nonumber\\
&+\frac{1}{\sqrt{N}}
\sum_{\mathbf{k}\sigma\sigma'}\sum_{i\neq j}
\eta_\mathbf{k}\,t_{ij}\,e^{-i\mathbf{kR}_i} 
\hat{f}_{i\sigma'}^\dagger\hat{f}_{i\sigma}^{\phantom{\dagger}}
c_{\mathbf{k}\sigma}^\dagger \hat{f}_{j\sigma'}^{\phantom{\dagger}}
+\text{h.~c.}
\nonumber\\
&+\frac{1}{N}\sum_{\mathbf{kk'}}\sum_{i\sigma}
e^{i(\mathbf{k}'-\mathbf{k})\mathbf{R}_i}\,
\eta_\mathbf{k} V_{\mathbf{k}'}
\,c_{\mathbf{k}\sigma}^\dagger
c_{\mathbf{k}'\sigma}^{\phantom{\dagger}}
P_i^0+\text{h.~c.}\nonumber\\
&-\frac{1}{N}\sum_{\mathbf{kk'}}\sum_{i\sigma\sigma'}
e^{i(\mathbf{k}'-\mathbf{k})\mathbf{R}_i} \,
\eta_\mathbf{k} V_{\mathbf{k}'}\, 
c_{\mathbf{k}\sigma}^\dagger 
c_{\mathbf{k}'\sigma'}^{\phantom{\dagger}}
\hat{f}_{i\sigma'}^\dagger\hat{f}_{i\sigma}^{\phantom{\dagger}}+\text{h.~c.}
\nonumber\\
&-\frac{1}{N}\sum_{\mathbf{k}\sigma}\sum_{ij}
2e^{-i\mathbf{k}(\mathbf{R}_i - \mathbf{R}_j)}
\eta_\mathbf{k} V_\mathbf{k}\,
\hat{f}_{i\sigma}^\dagger \hat{f}_{j\sigma}^{\phantom{\dagger}}
\end{align}
In the first term we recognise a contribution, which possesses the form
of the hybridisation interaction. The last term represents the
generated hopping of the correlated $f$ electrons, whose generation we have 
already discussed in the last paragraph. In terms of Hubbard operators
both contributions can be characterised as one-particle contributions,
whereas the remaining part is comprised of two-particle interactions.
The second and the third term both correspond to a hybridisation-like
interaction, which couples to an empty and a singly occupied $f$ site,
respectively. Alike the usual hybridisation term they alter the number
of particles in the respective electron subsystem by one. In contrast
the interactions in the fourth and fifth line do not change the number of
particles within the conduction band or within the $f$ electron
subsystem. The penultimate term can be regarded a Coqblin-Schrieffer
interaction. \par
To cope with all contributions of Eq.~\eqref{eq:commutator} we are obliged to introduce
a truncation scheme. The objective of this is to keep the Hamiltonian in the
form \eqref{eq:PAM_flow}. This task may be accomplished by various techniques,
one of them is the normal ordering of the operators \cite{Wegner1994}. Because
normal ordering must be defined with respect to a bilinear Hamiltonian, it is
more convenient to use an alternative scheme here. We simply decouple the
higher interactions by use of a Hartree-Fock decoupling. In this context
the fourth line of \eqref{eq:commutator} is evaluated as
\begin{align}\label{eq:decoupling}
&\frac{1}{N}\sum_{\mathbf{kk'}}\sum_{i\sigma}
e^{i(\mathbf{k}'-\mathbf{k})\mathbf{R}_i}
\,\eta_\mathbf{k} V_{\mathbf{k}'}
\,c_{\mathbf{k}\sigma}^\dagger
c_{\mathbf{k}'\sigma}^{\phantom{\dagger}}
P_i^0+\text{h.~c.}\nonumber\\
&\rightarrow\;2 \sum_{\mathbf{k}\sigma} \eta_\mathbf{k}
V_\mathbf{k}\langle P_i^0\rangle \,c_{\mathbf{k}\sigma}^\dagger
c_{\mathbf{k}\sigma}^{\phantom{\dagger}} -
\frac{2}{N} \sum_{i\mathbf{k}}\sum_{\sigma\sigma'} \eta_\mathbf{k}V_\mathbf{k} \langle
n_{\mathbf{k}\sigma}^c\rangle\,
\hat{f}_{i\sigma'}^\dagger
\hat{f}_{i\sigma'}^{\phantom{\dagger}}\nonumber\\
&+ \frac{2}{N} \sum_{i\mathbf{k}\sigma} \eta_\mathbf{k}V_\mathbf{k} \langle
n_{\mathbf{k}\sigma}^c\rangle \bigl(1 -\langle P_i^0\rangle\bigr)
\end{align}
where the representation $\langle n_{\mathbf{k}\sigma}^c\rangle\!=\!\langle
c_{\mathbf{k}\sigma}^\dagger c_{\mathbf{k}\sigma}^{\phantom{\dagger}}\rangle$
and the relation  $P_i^0+\sum_\sigma
\hat{f}_{i\sigma}^\dagger\hat{f}_{i\sigma}^{\phantom{\dagger}}=\mathbf{1}
$ were used. Here the expectation values are to be evaluated with regard to
the Hamiltonian \eqref{PAM}, thus they are constant during the entire flow.
Obviously, the first two terms on the right hand side 
contribute to the one-particle terms of the Hamiltonian, whereas the
last one results in an energy shift. It should be emphasised that this
factorisation scheme preserves the character of the initial
interaction with regard to the number of particles in each electron
subsystem. In Eq.~\eqref{eq:decoupling}, for example, both the interaction on
the left hand side and the resulting terms on the right hand side preserve the
number of $c$ and $f$ electrons. Furthermore, the local projector $P_i^0$
remains untouched by such a kind of approximation. A detailed derivation of the
decoupling of the remaining interaction is given in Appendix
\ref{sec:decoupling}. \par
If we collect all the terms that are provided by the factorisation
scheme, we recognise that the occurrence of the expectation value
$\langle P_i^0\rangle$ is always accompanied by the expectation value
$\langle n_{i\sigma}^f\rangle$. As we have mentioned above the first
denotes the average number of empty $f$ sites, and the latter
represents the number of singly occupied $f$ sites. Since we consider the
non-magnetic phase in the present paper, we can deduce the relation
$\langle n_{i\sigma}^f\rangle=(1-\langle P_i^0\rangle)/\nu_f$ from the
completeness relation for the local $f$ configuration space. As a
consequence of this we conclude that $\langle P_i^0\rangle$ is of
order $\nu_f^0$ whereas $\langle n_{i\sigma}^f\rangle$ is of order
$\nu_f^{-1}$. The same becomes obvious for the contributions for the
energy $\varepsilon_f$, where the terms of order $\nu_f^0$ come from
the parts $\propto P_i^0$, and the terms of order $\nu_f^{-1}$ are given
by the contributions $\propto \hat{f}_{i\sigma}^\dagger
\hat{f}_{i\sigma}^{\phantom{\dagger}}$. As a simplification we
consider in the following the case of great degeneracy of the
electrons, i.e. $\nu_f\rightarrow\infty$, and neglect all terms of
order $\nu_f^{-1}$. The spirit of this approximation is rather similar
to the SBMF theory, though both approaches cannot be simply
compared on this level. Eventually, we arrive at the following system
of differential equations for the one-particle energies of the
Hamiltonian \eqref{eq:PAM_flow}
\begin{align}
\label{DGL:ek1}
\frac{\mathrm{d}\varepsilon_\mathbf{k}}{\mathrm{d}\ell}=&
2\langle P_i^0\rangle \eta_\mathbf{k} V_\mathbf{k}\\ 
\label{DGL:ef1}
\frac{\mathrm{d}\Delta_\mathbf{k}}{\mathrm{d}\ell}=&
-2 \eta_\mathbf{k} V_\mathbf{k}\\
\label{DGL:Ef}
\frac{\mathrm{d}\varepsilon_f}{\mathrm{d}\ell}=&
-\frac{2}{N}\sum_{\mathbf{k}\sigma}\langle n_{\mathbf{k}\sigma}^c\rangle
\eta_\mathbf{k} V_\mathbf{k}
- \frac{1}{N}\sum_{\mathbf{k}\sigma} \langle
A_{\mathbf{k}\sigma}\rangle  \eta_\mathbf{k} \Delta_\mathbf{k}
\\
\label{DGL:V}
\frac{\mathrm{d}V_\mathbf{k}}{\mathrm{d}\ell}=&
-\bigl(\varepsilon_\mathbf{k}
-\varepsilon_f-\langle 
P_i^0\rangle\Delta_\mathbf{k} \bigr) \eta_\mathbf{k}
\\
\label{DGL:E}
\frac{\mathrm{d}E}{\mathrm{d}\ell}=&\frac{2}{N}\sum_{i\mathbf{k}\sigma}
\langle n_{\mathbf{k}\sigma}^c\rangle\bigl(1- \langle P_i^0\rangle 
\bigr) \eta_\mathbf{k}V_\mathbf{k} \nonumber\\
&+\frac{1}{N}\sum_{i\mathbf{k}\sigma}\langle
A_{\mathbf{k}\sigma} \rangle \bigl(1-\langle P_i^0\rangle \bigr) 
\eta_\mathbf{k}\Delta_\mathbf{k}
\end{align}
The system possesses the initial conditions
$\varepsilon_\mathbf{k}(\ell=0)=\varepsilon_\mathbf{k}^\mathrm{i}$,
$\varepsilon_f(\ell=0)=\varepsilon_f^\mathrm{i}$,
$V_\mathbf{k}(\ell=0)=V_\mathbf{k}^\mathrm{i}$ and
$\Delta_\mathbf{k}(\ell=0)=E(\ell=0)=0$. From the above equations we
can immediately conclude $\langle P_i^0\rangle
\Delta_\mathbf{k}(\ell)=\varepsilon_\mathbf{k}^\mathrm{i}-\varepsilon_\mathbf{k}(\ell)$
and $E(\ell)=N(\varepsilon_f^\mathrm{i}-\varepsilon_f(\ell))(1-\langle
P_i^0\rangle)$ so that $\Delta_\mathbf{k}$ can be eliminated completely,
by making use of the first relation.\par
At this point we emphasise that we have not used perturbational
arguments with respect to the hybridisation strength nor have we used a
mean-field decoupling of the 
Coulomb interaction within the above approximation scheme. This fact is very
important with regard to the description of heavy fermion behaviour.\par
With the form-invariant Hamiltonian \eqref{eq:PAM_flow} we know that in the
limit $\ell\rightarrow\infty$ the
resulting effective Hamiltonian  is of the form
\begin{align}\label{eq:diagonal}
  H_\mathrm{PAM}^\ast=\sum_{\mathbf{k}\sigma} \varepsilon_\mathbf{k}^\ast
  c_{\mathbf{k}\sigma}^\dagger c_{\mathbf{k}\sigma}^{\phantom{\dagger}}
  + \sum_{\mathbf{k}\sigma} \bigl( \varepsilon_f^\ast +
  \Delta_\mathbf{k}^\ast\bigr)
  \hat{f}_{\mathbf{k}\sigma}^\dagger
  \hat{f}_{\mathbf{k}\sigma}^{\phantom{\dagger}}
  + E^\ast
\end{align}
It describes two decoupled subsystems of uncorrelated and correlated electrons
and can therefore not be regarded as diagonal. This fact is the main
difference to other approaches, such as the SBMF theory, which leave
a noninteracting Hamiltonian. Another important difference is 
that we do not need to incorporate a side condition, which restricts
the number of particles on each $f$ sites, for we did not made use of
auxiliary boson fields.
\subsection{Flow of Observables}\label{sec:observ}
In the introduction we have already discussed the representation of
expectation values within the framework of flow equations. There we concluded
that expectation values as well as correlation functions can be easily
evaluated on the basis of the derived effective Hamiltonian, if we regard the
transformed operators at the same time. The transformation of an arbitrary
operator works quite similar to the transformation of the Hamiltonian. If we
consider, in the first instance, the flow of the operator
$c_{\mathbf{k}\sigma}$, we can write the mapping in the form
\begin{align}\label{eq:c}
  \frac{\mathrm{d} c_{\mathbf{k}\sigma}}{\mathrm{d}\ell}=\bigl[
    \eta,c_{\mathbf{k}\sigma}\bigr]
\end{align}
Similar to the transformation of the Hamiltonian we have to regard all the
contributions that are
generated by the commutator on the right hand side. The operator, of course,
has to fulfil the boundary condition
$c_{\mathbf{k}\sigma}(0)=c_{\mathbf{k}\sigma}$. Hence we obtain the lowest
order contributions simply by evaluating the commutator with
$c_{\mathbf{k}\sigma}$. With this results we can infer the following form for
the $\ell$-dependent operator
\begin{equation}
  c_{\mathbf{k}\sigma}(\ell)=\alpha_\mathbf{k}\, c_{\mathbf{k}\sigma} 
  + \beta_\mathbf{k}\, \hat{f}_{\mathbf{k}\sigma}
\end{equation}
This is a linear combination of the ordinary operators $c_{\mathbf{k}\sigma}$
and $\hat{f}_{\mathbf{k}\sigma}$ with the weighting factors $\alpha_\mathbf{k}$
and $\beta_\mathbf{k}$, respectively. It is the simplest
possible representation for the flow of the operator $c_{\mathbf{k}\sigma}$,
and higher interactions that are generated by the commutator must be
truncated in the same manner as we  have treated the flow of the
Hamiltonian \cite{nuf}. The complete evaluation of the flow equation
\eqref{eq:c} yields  differential equations for the weighting factors
\begin{equation}
  \frac{\mathrm{d}\alpha_\mathbf{k}}{\mathrm{d}\ell}=
  -\beta_\mathbf{k}\eta_\mathbf{k} \langle P_i^0\rangle\;,\quad 
  \frac{\mathrm{d}\beta_\mathbf{k}}{\mathrm{d}\ell}=\alpha_\mathbf{k}\eta_\mathbf{k}
\end{equation}
Here again the $\ell$-dependence has been omitted for the reason of
readability. Obviously, the initial values of the weight factors are given by
$\alpha_\mathbf{k}^\mathrm{i}=1$ and $\beta_\mathbf{k}^\mathrm{i}=0$, and as a
consequence of unitarity they obey the relation 
\begin{equation}\label{eq:unitarity}
  \alpha_\mathbf{k}^2 + \langle P_i^0\rangle \beta_\mathbf{k}^2=1
\end{equation}
In a further step the coefficient $\beta_\mathbf{k}$ can be eliminated, and
the flow of $\alpha_\mathbf{k}$ is governed by the equation
\begin{equation}\label{DGL:alpha}
  \frac{\mathrm{d}}{\mathrm{d}\ell}\arcsin \bigl[1-2\alpha_\mathbf{k}^2\bigr]
  =2\sqrt{\langle P_i^0\rangle} \eta_\mathbf{k}
\end{equation}
The transformation of the operator $\hat{f}_{\mathbf{k}\sigma}$ can be treated
likewise, and we obtain
\begin{equation}
  \hat{f}_{\mathbf{k}\sigma}(\ell)=-\langle P_i^0\rangle \beta_\mathbf{k}\,  c_{\mathbf{k}\sigma}
  + \alpha_\mathbf{k}\, \hat{f}_{\mathbf{k}\sigma}
\end{equation}
With the results of the previous paragraph we are now in the position
to evaluate all expectation values that occur in the differential equations by
using the unitarity of the transformation, i.~e. $\langle
n_{\mathbf{k}\sigma}\rangle = {}^\ast\langle
c_{\mathbf{k}\sigma}^{\ast\dagger}
c_{\mathbf{k}\sigma}^\ast\rangle^\ast$. Here
${}^\ast\langle\cdots\rangle^\ast$ refers to the expectation value with
respect to the fully transformed Hamiltonian \eqref{eq:diagonal}. By inserting the
transformed operators, we obtain for the non-local expectation values 
\begin{equation}\label{eq:nk}
  \langle n_{\mathbf{k}\sigma}\rangle 
  = ( \alpha_\mathbf{k}^\ast)^2\, {}^\ast \langle
  c_{\mathbf{k}\sigma}^\dagger c_{\mathbf{k}\sigma}^{\phantom{\dagger}}\rangle^\ast
  + (\beta_\mathbf{k}^\ast)^2\, {}^\ast \langle
  \hat{f}_{\mathbf{k}\sigma}^\dagger
  \hat{f}_{\mathbf{k}\sigma}^{\phantom{\dagger}}\rangle^\ast
\end{equation}
and
\begin{equation}\label{eq:Ak}
  \langle A_{\mathbf{k}\sigma}\rangle =
  2\,\alpha_\mathbf{k}^\ast\beta_\mathbf{k}^\ast\, \bigl( 
  -\langle P_i^0 \rangle {}^\ast\langle c_{\mathbf{k}\sigma}^\dagger
  c_{\mathbf{k}\sigma}^{\phantom{\dagger}} \rangle^\ast + 
  {}^\ast\langle \hat{f}_{\mathbf{k}\sigma}^\dagger
  \hat{f}_{\mathbf{k}\sigma}^{\phantom{\dagger}} \rangle^\ast \bigr)
\end{equation}
In view of the Hamiltonian \eqref{eq:diagonal} it is clear that there are no
contributions from expectation values of the form ${}^\ast\langle
c_{\mathbf{k}\sigma}^\dagger
\hat{f}_{\mathbf{k}\sigma}^{\phantom{\dagger}}\rangle^\ast$.\par
The derivation of the non-local expectation values, which contain contributions
from different lattice sites, could be carried out rather intuitively, since we
made use of a descriptive transformation scheme for the single operators. In
contrast, the treatment of the local expectation value $\langle P_i^0\rangle$
turns out to be more difficult. Though the projector can be represented in
terms of Hubbard operators as
$P_i^0=\hat{f}^\dagger_{i\sigma}\hat{f}_{i\sigma}^{\phantom{\dagger}}$, its 
transformation cannot simply be carried out by the transformation of the
single operators, for this procedure would break the restriction for the
$f$ electrons to empty and
singly occupied sites. Instead of that we must transform the entire operator
$P_i^0$. The detailed derivation of this expectation value is presented in
Appendix \ref{sec:local_expectation}, and as result we obtain
\begin{align}\label{eq:P}
\langle P_i^0 \rangle= 1-\frac{1}{N}\sum_{\mathbf{k}\sigma}
\biggl[&\bigl(1-(\alpha_\mathbf{k}^\ast)^2\bigr) \;{}^\ast\langle  
c_{\mathbf{k}\sigma}^\dagger
c_{\mathbf{k}\sigma}^{\phantom{\dagger}}\rangle^\ast\nonumber\\
&+
\bigl(1-(\beta_\mathbf{k}^\ast)^2\bigr)\;
{}^\ast\langle \hat{f}_{\mathbf{k}\sigma}^\dagger
\hat{f}_{\mathbf{k}\sigma}^{\phantom{\dagger}}\rangle^\ast
\biggr]
\end{align} 
Altogether, this more sophisticated treatment can also be applied in order to
derive the non-local expectation values, if we transform them as pairs rather
than as single operators. The result of this evaluation, though,
coincides with the above derivation, and we can use the much easier
representation from the last paragraph first for reasons of clearness and
second as a starting point for the evaluation of the Green's functions.\par
Further on, we wish to discuss the form of the expectation value which is
provided by \eqref{eq:P}. If we alternatively start out from the fact that
the total number of particles must be conserved under the unitary
transformation, we can derive this result in another way. The relation 
\begin{equation}
  \sum_{\mathbf{k}\sigma} \langle
  n_{\mathbf{k}\sigma}^c\rangle + \sum_{i\sigma}\langle
  \hat{f}_{i\sigma}^\dagger \hat{f}_{i\sigma}^{\phantom{\dagger}}
  \rangle = \sum_{\mathbf{k}\sigma} {}^\ast\langle
  c_{\mathbf{k}\sigma}^\dagger
  c_{\mathbf{k}\sigma}^{\phantom{\dagger}}\rangle^\ast
  + \sum_{\mathbf{k}\sigma} {}^\ast\langle
  \hat{f}_{\mathbf{k}\sigma}^\dagger
  \hat{f}_{\mathbf{k}\sigma}^{\phantom{\dagger}} \rangle^\ast
\end{equation}
denotes the particle number conservation. If we use on the left hand side the
expression \eqref{eq:nk} for the expectation value $\langle
n_{\mathbf{k}\sigma}^c\rangle$ and divide by the number of sites,  we
eventually arrive at Eq:~\eqref{eq:P}. Consequently, we can deduce that the
above derivation is 
consistent with the conservation of the total number of particles.\par
As a remaining task we have to evaluate the expectation values ${}^\ast\langle
c_{\mathbf{k}\sigma}^\dagger
c_{\mathbf{k}\sigma}^{\phantom{\dagger}}\rangle^\ast$ and $ {}^\ast\langle
\hat{f}_{\mathbf{k}\sigma}^\dagger 
\hat{f}_{\mathbf{k}\sigma}^{\phantom{\dagger}}\rangle^\ast$ with respect to
the effective Hamiltonian \eqref{eq:diagonal}. As the subsystem of
$c$ electrons simply consists of non-interacting fermions the respective
expression 
is determined by a Fermi distribution. The calculation of the remaining
$f$-subsystem is more complex as these particles do not obey a simple
statistics. It is expedient to calculate the demanded expectation
value by means of ordinary  Green's function calculations
\cite{Abrikosov1963}. The equation of motion for the $f$ Green's
function is simply given by
\begin{align}
\omega\;\langle\langle
\hat{f}_{\mathbf{k}\sigma}^{\phantom{\dagger}};
\hat{f}_{\mathbf{k}\sigma}^\dagger\rangle\rangle(\omega+i\delta)   
=& \bigl\langle\bigl\{ 
\hat{f}_{\mathbf{k}\sigma}^{\phantom{\dagger}},\hat{f}_{\mathbf{k}\sigma}^\dagger
\bigr\}_+\bigr\rangle\nonumber\\
&+ \bigl\langle\bigl\langle
\bigl[\hat{f}_{\mathbf{k}\sigma},H^\ast\bigr];
\hat{f}_{\mathbf{k}\sigma}^\dagger\bigr\rangle\bigr\rangle
(\omega+i\delta)
\end{align}
where we set $\hbar=1$.
This representation suggests a coupling to higher order Green's functions, which
can be seen by the second term on the right hand side. In order to
obtain a closed equation of motion we again apply the decoupling scheme of
the last section and neglect contributions of order $\nu_f^{-1}$.  As result
we obtain 
\begin{equation}\label{eq:korrelation_f}
\langle\langle \hat{f}_{\mathbf{k}\sigma}^{\phantom{\dagger}};
\hat{f}_{\mathbf{k}\sigma}^\dagger\rangle\rangle(\omega+i\delta) =
\frac{\langle P_i^0\rangle}{  
\omega + is - (\varepsilon_f^\ast +\langle P_i^0\rangle
\Delta_\mathbf{k}^\ast)}
\end{equation}
and, by the application of the spectral theorem, the desired
expectation value reads 
\begin{equation}\label{eq:correlation_f}
 {}^\ast\langle\hat{f}_{\mathbf{k}\sigma}^\dagger
\hat{f}_{\mathbf{k}\sigma}^{\phantom{\dagger}}\rangle^\ast
 =\langle
P_i^0 \rangle\; n_\mathrm{F}\bigl(\varepsilon_f^\ast + \langle P_i^0 \rangle
\Delta_\mathbf{k}^\ast\bigr)
\end{equation}  
On the right hand side the Fermi distribution
$n_\mathrm{F}=(1+\exp(\beta E))^{-1}$ has been introduced.
\section{Integration of the flow equations}
\subsection{Analytical solution}\label{sec:ana}
The PAM during the flow is given by Eq.~\eqref{eq:PAM_flow}
and leads to an effective Hamiltonian of the form \eqref{eq:diagonal},
provided we have chosen a generator which ensures the vanishing of the
hybridisation $V_\mathbf{k}$. The evolution of the matrix elements is
given by a closed set of coupled differential equations, which have to
be evaluated for the derivation of physical quantities. In a first
step we try to find an analytical solution. For this purpose we bring
Eq. \eqref{DGL:V} in the following form
\begin{equation}\label{eq:eta_k}
  \eta_\mathbf{k} = -\frac{1}{2\varepsilon_\mathbf{k}
  -\varepsilon_f
  -\varepsilon_\mathbf{k}^\mathrm{i}}\frac{\mathrm{d}V_\mathbf{k}}
{\mathrm{d}\ell}
\end{equation}
Here we have substituted $\Delta_\mathbf{k}$ following the above
discussion. This equation provides a representation of
$\eta_\mathbf{k}$ as function of the one-particle energies and allows us 
to eliminate this quantity in all other differential equations.  
With Eq. \eqref{eq:eta_k} we are now in the position to rewrite the differential
equation for the single-particle energy $\varepsilon_\mathbf{k}$ as
\begin{equation}\label{DGL:ek2}
\frac{\mathrm{d}}{\mathrm{d}\ell} \bigl[
\varepsilon_\mathbf{k}^2 - \varepsilon_\mathbf{k}(\varepsilon_f^\ast
+\varepsilon_\mathbf{k}^\mathrm{i}) + \langle P_i^0\rangle V_\mathbf{k}^2
\bigr]= (\varepsilon_f
-\varepsilon_f^\ast)\frac{\mathrm{d}\varepsilon_\mathbf{k}}
{\mathrm{d}\ell}
\end{equation}
As a simplification for the next discussions we assume that the energy
$\varepsilon_f$ converges rather fast to its limit value, while
the main contribution to the evolution of the other energies occur at
a larger $\ell$-scale. This assumption  allows us to replace
$\varepsilon_f(\ell)$ by the constant energy $\varepsilon_f^\ast$. Such kind
of simplification is quite common in terms of flow equations 
\cite{Kehrein1996,Kehrein1996a}.  
With this approximation the equation \eqref{DGL:ek2} can simply be integrated
and we arrive at
\begin{equation}\label{GL_V2}
 (\varepsilon_\mathbf{k}^\ast)^2 
-\varepsilon_\mathbf{k}^\ast\bigl(\varepsilon_f^\ast+
\varepsilon_\mathbf{k}^\mathrm{i} \bigr)
+\varepsilon_\mathbf{k}^\mathrm{i} \varepsilon_f^\ast-\langle P_i^0\rangle
(V_\mathbf{k}^\mathrm{i})^2=0 
\end{equation}
using the fact that $V_\mathbf{k}$ vanishes in the limit
$\ell\rightarrow\infty$. This quadratic equation has the solutions
\begin{equation}\label{eq:ana1}
\varepsilon_\mathbf{k}^\ast= \frac{1}{2}
\bigl(\varepsilon_f^\ast+
\varepsilon_\mathbf{k}^\mathrm{i}\bigr)\pm \frac{1}{2}W_\mathbf{k}
\end{equation}
where we have used the representation 
\begin{equation}
W_\mathbf{k}=\sqrt{\bigl(\varepsilon_f^\ast-
\varepsilon_\mathbf{k}^\mathrm{i}\bigr)^2 + 4\langle P_i^0\rangle
(V_\mathbf{k}^\mathrm{i})^2}
\end{equation}
Equivalently, we obtain the following solution for the $f$ degrees of freedom
\begin{equation}\label{eq:anad}
  \varepsilon_f^\ast + \langle P_i^0 \rangle
  \Delta_\mathbf{k}^\ast=\frac{1}{2} 
\bigl(\varepsilon_f^\ast+
\varepsilon_\mathbf{k}^\mathrm{i}\bigr)\mp \frac{1}{2}W_\mathbf{k}
\end{equation}
The latter describes the excitation energies of the $f$ electrons as suggested
by Eq.~\eqref{eq:correlation_f}. In order to derive a simple equation for the
energy $\varepsilon_f$ we neglect the term with $\langle
A_\mathbf{k\sigma}\rangle$ on the right hand side of \eqref{DGL:Ef}, and after
having integrated the resulting equation, we obtain 
\begin{equation}\label{eq:ana2}
\varepsilon_f^\ast
 -\varepsilon_f^\mathrm{i}=-\frac{1}{N}\sum_{\mathbf{k}\sigma}
 \frac{\langle 
 n_{\mathbf{k}\sigma}^c\rangle}{\langle P_i^0\rangle} (\varepsilon_\mathbf{k}^\ast
 -\varepsilon_\mathbf{k}^\mathrm{i})=
-\frac{1}{N}
\sum_{\mathbf{k}\sigma}
\frac{\langle n_{\mathbf{k}\sigma}^c\rangle (V_\mathbf{k}^\mathrm{i})^2}
{\varepsilon^\ast_f-\varepsilon^\ast_\mathbf{k}}
\end{equation}
The simplification of a constant  energy $\varepsilon_f$ can also be made for
the derivation of the weighting factors and allow us the evaluation of
expectation values. From \eqref{DGL:alpha} we arrive at 
\begin{equation}
(\alpha_\mathbf{k}^\ast)^2=\frac{1}{2}\biggl( 1+
\frac{\varepsilon_f^\ast-\varepsilon_\mathbf{k}^\mathrm{i}}{W_\mathbf{k}}\biggr)
\end{equation}
Altogether the results coincide with the expressions from the SBMF
theory. The excitation energies \eqref{eq:ana1} and \eqref{eq:anad}
constitute the electronic structure of the system of two gapped
bands. The overall renormalisation of the $f$ level is provided by
Eq. \eqref{eq:ana2}. These results are based on the assumption of a
fast convergence of $\varepsilon_f$ compared to all other energies and
give us a rough estimate of the electronic structure of the full
problem. However, a thorough examination beyond this simplification requires a
numerical treatment of the differential equations. We would like to stress 
that we have not used a particular choice for the coefficient
$\eta_\mathbf{k}$. Thus the derived results are valid for any
$\eta_\mathbf{k}$ that yields a vanishing hybridisation for the
fully transformed Hamiltonian.
\subsection{Numerical solution}
The above derivation of an analytical solution for the differential equations
was based on further simplifications. A more general solution can be provided
by a numerical integration. While the analytical solution was
independent of a particular choice of the coefficient $\eta_\mathbf{k}$, 
we now need to determine this quantity for the numerical evaluation.
An expedient choice is given by
\begin{equation}\label{eq:num_gen}
  \eta_\mathbf{k}= \bigl(2
  \varepsilon_\mathbf{k} -\varepsilon_f -
  \varepsilon_\mathbf{k}^\mathrm{i}\bigr)\, V_\mathbf{k}
\end{equation}
since the energy $V_\mathbf{k}$ vanishes when $\ell$ goes to infinity. This
becomes obvious by inserting Eq. \eqref{eq:num_gen} into Eq. \eqref{DGL:V},
for the differential of $V_\mathbf{k}$ is a negative quantity in the entire
integration range, while $V_\mathbf{k}$ itself is always positive. The above 
choice is not dissimilar to Wegner's generator. However, it is not simply
obtained by commuting the interacting and the non-interacting part of the
Hamiltonian, as suggested by Eq. \eqref{eq:wegner}, in that the non-interacting
part for the present calculations 
incorporates the hopping of strongly correlated objects. In fact, the
derivation of the generator \eqref{eq:num_gen} uses a factorisation
scheme, which has been presented in detail in Sec.~\ref{sec:flow}. This
particular form of the generator also triggers the Hamiltonian flowing
into a stable fixed point, as $\eta$ commutes with $H_\mathrm{PAM}$ in
the limit $\ell\rightarrow\infty$. As a consequence the differentials
of the one-particle energies vanish in this limit, which can be seen on
the basis of Eqs.~\eqref{DGL:ek1}-\eqref{DGL:E}.\par
The numerics generally comprises two tasks, one of them being the numerical
integration of the flow equations, which is done by a Runge-Kutta
integration of fifth order with adaptive step size control. The second
part of the numerical calculation is the self-consistent
evaluation of the expectation values $\langle P_i^0\rangle$, $\langle
n_{\mathbf{k}\sigma}^c\rangle$ and $\langle
A_{\mathbf{k}\sigma}\rangle$.\par
Since the differential equations are coupled and possess a dependence
on $\mathbf{k}$-dependent expectation values, the numerical
integration turns out to be a rather arduous task. Consequently, we
restrict ourselves to the solution of a one-dimensional
system. As we consider merely the paramagnetic phase of the
PAM, we believe that one dimension is sufficient to examine the most
general aspects of heavy-fermion behaviour.
\section{Results}\label{sec:results}
In the following section we present the results of the numerical calculations for
the PAM. In a first step we analyse the flow of the one-particle
energies of the Hamiltonian. All
calculations are performed for systems of $1000$ sites in one dimension. As we
have mentioned above, we merely consider a dispersion-less
hybridisation. Notwithstanding, the  hybridisation $V_\mathbf{k}$ shows  distinct
flows for different wave vectors $\mathbf{k}$ due to its dependence on
the electron energies (cf. Eq.~\eqref{DGL:V}). For the conduction
electrons we use a constant density of states (DOS), i.~e. a linear
dispersion in one dimension, and we consider $T=0$.\par
For the upcoming discussions we use lattice constant of $a=1$ and
$W=2$ for the bandwidth of the bare conduction band, so that all 
energies are given in units of the half band width, which is usually
of order $1-10~\mathrm{eV}\sim 10^{4}-10^{5}~\mathrm{K}$. As we
consider one-dimensional systems, the wave vector is rather referred to
as wave number. 
\subsection{One-particle energies}\label{sec:one-particle}
In the following discussion we address the
one-particle energies $\varepsilon_\mathbf{k}$, $\varepsilon_f$ and
$V_\mathbf{k}$. For a particular parameter regime their flow diagrams
are
\begin{figure}
  \includegraphics[width=\linewidth]{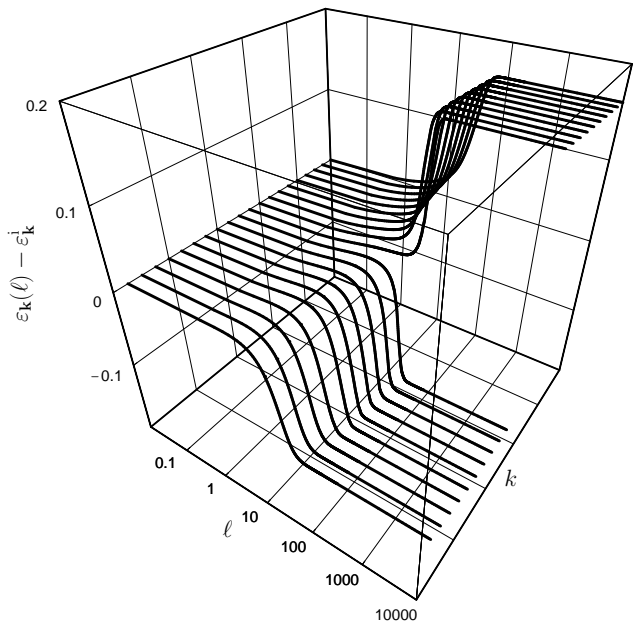}
  \caption{\label{fig:ek3d}Flow of the quantities
  $\varepsilon_\mathbf{k}-\varepsilon_\mathbf{k}^\mathrm{i}$ for subset wave
    numbers near the critical wave number (see text) for a system with
  $\varepsilon_f^\mathrm{i}=0.9$, $V^\mathrm{i}=0.3$ and $\nu_f=2$.}
\end{figure}
\begin{figure}
  \includegraphics[width=\linewidth]{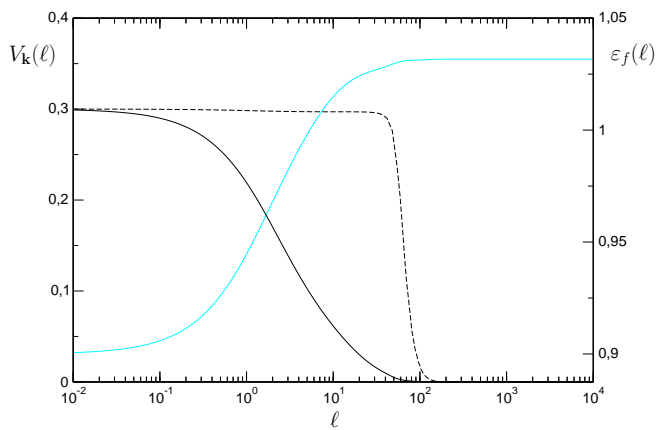}
  \caption{\label{fig:EfV_flow}Flow of the quantities 
  $V=1/N\sum_\mathbf{k}V_\mathbf{k}$ (black solid line, left scale),
  $V_\mathbf{k}$ for the critical wave number (black dashed line, left scale)
  and $\varepsilon_f$ (blue solid line, right scale). The system parameters
  are as in Fig.~\ref{fig:ek3d}.}
\end{figure}
depicted in Fig.~\ref{fig:ek3d} and Fig.~\ref{fig:EfV_flow}.
For the electron energies it is 
more convenient to investigate the difference to their initial value,
namely $\varepsilon_\mathbf{k}(\ell)-\varepsilon_\mathbf{k}^\mathrm{i}$, as a
function of the flow parameter $\ell$. This quantity is shown in
Fig.~\ref{fig:ek3d} for a range of wave numbers near the crossing point. The
latter is characterised as the particular wave number, that divides the
energies between increasing and decreasing quantities. This behaviour can be 
understood on the basis of Eq.~\eqref{DGL:ek1}, as the sign of the
right hand side depends on the wave number. For the system in
Fig.~\ref{fig:ek3d} we find positive energy differences
$\varepsilon_\mathbf{k}^\ast-\varepsilon_\mathbf{k}^\mathrm{i}$ above a
critical wave number of $\mathbf{k}=0.499\pi$. As the graph shows, all
energies start off with small modifications up to values of
$\ell\simeq 1$ before they enter a region of significant rise or fall. In the region
of $\ell\gtrsim 10^2$ they have already converged to their limit
value. The most remarkable result of their behaviour is definitely the sharp
crossover between increasing and decreasing energies. \par
A much simpler picture is provided by the graphs in Fig.~\ref{fig:EfV_flow}. As demanded,
the hybridisation energy $V_\mathbf{k}$ vanishes for all wave vectors for
$\ell\gtrsim 10^2$. In Fig.~\ref{fig:EfV_flow} the flow of the quantity
$V=1/N\sum_\mathbf{k} V_\mathbf{k}$ is shown (solid black line) beside the
flow of the hybridisation $V_\mathbf{k}$ for the critical wave number
$\mathbf{k}=0.499\pi$. For the latter the onset of a significant
modification is delayed to higher values of $\ell$, and it drops sharply to
zero in the region of $\ell\simeq 10^2$. This suggests that the
$V_\mathbf{k}$ with wave numbers near the upper and lower band edge are 
renormalised at lower values of $\ell$. Altogether we can conclude 
that all hybridisation elements can be considered zero above
$\ell\simeq10^2$ . On the other hand the one-particle energy which  
characterises the local $f$ level sees a sharp increase in the region of
$\ell\sim 10^0-10^1$ and converges rather fast. The numerical results
show that the present choice for $\eta_\mathbf{k}$ triggers a smooth
and continuous evolution of the one-particle energies of the
Hamiltonian \eqref{eq:PAM_flow}. The convergence to their limit
values occurs on a characteristic scale in the region of $\ell\gtrsim
10^2$. For the present results we use the bound $V=10^{-10}$ as
convergence criteria for the numerical integration of the flow
equations. This corresponds to a 
value of the flow parameter $\ell\sim 3000-5000$ in the considered 
parameter regimes. However, for a better  
clearness the integration for the system in Fig.~\ref{fig:EfV_flow} is
carried out up to a value of $\ell=10^4$.\par 
On the one hand we have a rather good convergence of the flow
equations due to their smoothness and their evolution to a stable
fixed point. On the other hand the self-consistent evaluation of the
expectation values is restricted to certain parameter regimes. From
the SBMF theory we know that it cannot describe the phase where all
$f$ sites are singly occupied, i.~e. $\langle
P_i^0\rangle=0$. Likewise, in the present approach we only obtain
valid results for a finite number of empty $f$ sites, and the solution
breaks down as $\langle 
P_i^0\rangle$ vanishes. This breakdown coincides with the energy
$\varepsilon_f^\ast$ dropping below the chemical potential. However, it is
possible to obtain solutions near this critical point. The limit of singly
occupied $f$ sites can be approached either by decreasing the hybridisation
energy or by increasing the difference $\mu-\varepsilon_f^\mathrm{i}$. With a
fixed value for $\varepsilon_f^\mathrm{i}$ the solution breaks down below a
critical value of $V^\mathbf{i}$, and, similarly, this point is reached for a
fixed $V^\mathrm{i}$ below a critical $\varepsilon_f^\mathrm{i}$. Stable 
solutions for small expectation values 
$\langle P_i^0\rangle$ can be obtained, preferably, at rather moderate values
of $\mu-\varepsilon_f^\mathrm{i}$ and small $V^\mathrm{i}$. This restriction is
believed to be lifted up by considering corrections of order $\nu_f^{-1}$
\cite{Fulde1993}. \par
Within the above factorisation scheme it is not sufficient to
incorporate the $\nu_f^{-1}$ contributions simply by considering them in
the Eqs.~\eqref{DGL:ek1}-\eqref{DGL:E}. Indeed, it turns out that this
procedure would lead to unphysical solutions escpecially in the case of small
$\nu_f$. On the other hand the dynamics of the $f$ electrons is even more
affected by these corrections and cannot be properly calculated in the limit
of small degeneracy. A consistent consideration of the $\nu_f$ therefore
needs an enhancement of the factorisation process. As we pointed out in
Sec.~\ref{sec:flow} the decoupling of the higher order interactions
preserves their character with respect to the variation of the particle
numbers for each subsystem. This assumption is too restrictive in the case of
small  degeneracy, as charge fluctuations which are induced by the flow of the
Hamiltonian become more important. Obviously, the extension of the
factorisation scheme in this sense would lead to a more complex system
of differential equations and is not considered in the present work.
\subsection{Electronic properties}
\begin{figure}[h]
  \includegraphics[width=\linewidth]{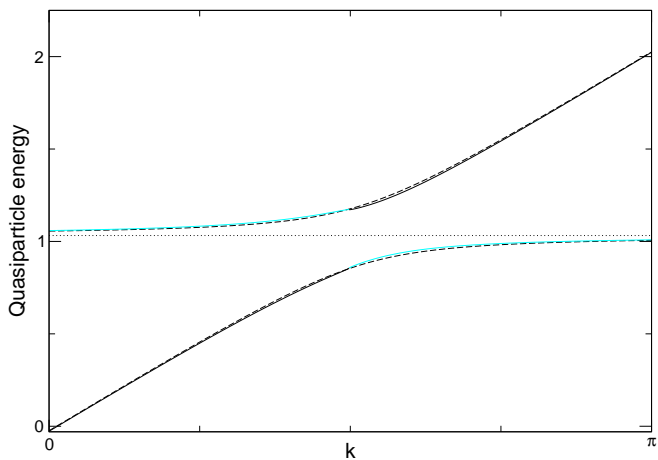}
  \caption{\label{fig:dispersionEf9V3} Dispersion of the quasiparticles for
  the system with parameters as in Fig.~\ref{fig:ek3d}. The
  energies $\varepsilon_\mathbf{k}^\ast$ and 
  $\varepsilon_f^\ast + \langle P_i^0\rangle \Delta_\mathbf{k}^\ast$ are
  represented by solid black and blue lines, respectively, whereas the the
  analytical dispersions are represented by staggered lines. The dotted line
  stands for $\varepsilon_f^\ast$.}
\end{figure}

\begin{figure}[h]
  \includegraphics[width=\linewidth]{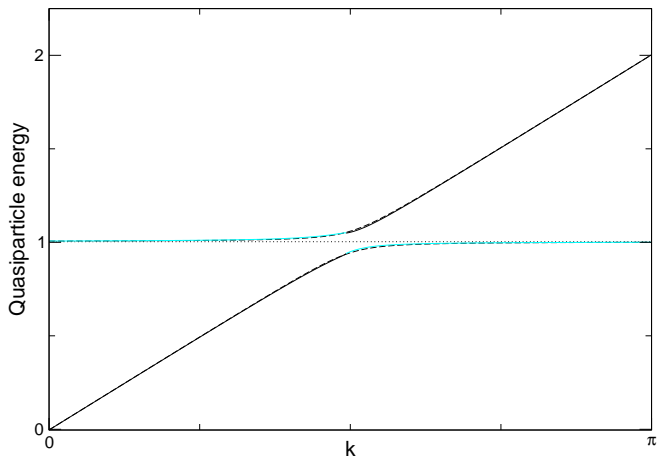}
  \caption{\label{fig:dispersionEf9V2} Dispersion of the quasiparticles as in
  Fig. \ref{fig:dispersionEf9V3} for a system with
  $\varepsilon_f^\mathrm{i}\!=\!0.9$, $V^\mathrm{i}\!=\!0.2$ and 
  $\nu_f\!=\!2$.}
\end{figure}

In this section we present the numerical results for the electronic
structure of the PAM. 
The dispersion relations of the quasiparticle energies are shown in
Fig.~\ref{fig:dispersionEf9V3} and Fig.~\ref{fig:dispersionEf9V2}.

The general form of the one-particle energies is  
quite similar to the results of the SBMF theory. In both approaches the
spectrum splits in two bands which are divided by a finite gap. Considering the 
chemical potential lying in one of the bands, the system possesses a
metallic ground state. The situation of an insulator is also
conceivable, if $\mu$ is right in the gap. In the following
considerations we only focus on the metallic 
case, though the insulating case can also be examined within the framework of
flow equations \cite{Meyer2003}.\par
The quasiparticle energies in Fig.~\ref{fig:dispersionEf9V3} are obtained for a
system with $\varepsilon_f^\mathrm{i}=0.9$, $V^\mathrm{i}=0.3$ and
$\nu_f=2$. The chemical potential is at $\mu=1$ and the bandwidth of the conduction
electron band is $W=2$. Since we have neglected terms of order $\nu_f^{-1}$ within our
calculations, the presented results for a twofold degenerate $f$ level are
deemed an extrapolation down to small values of $\nu_f$. This kind
of assumption has been frequently used in terms of various $1/\nu_f$
expansions in order to describe physical relevant systems. The excitation
energies $\varepsilon_\mathbf{k}^\ast$ (solid black line) and
$\varepsilon_f^\ast + \langle P_i^0\rangle$ (solid blue line) form two
continuous bands, that are separated by an indirect gap. The latter denotation
refers to the fact, that the minimal excitation energy from the lower to the
upper band occurs at different wave vectors. The renormalised energy
$\varepsilon_f^\ast$ (dotted line) lies right in the centre of this gap. The
self-consistently evaluated results for this energy and for the expectation
value of the empty $f$ 
sites are obtained as $\varepsilon_f^\ast=1.032$ and $\langle
P_i^0\rangle=0.285$, which states, that the system is well within the
mixed-valent regime. Though the results from the SBMF theory
of $\varepsilon_{f,\mathrm{SB}}^\ast=1.111$ and $\langle
P_i^0\rangle_\mathrm{SB}=0.577$ also suggest a mixed-valent 
system, the latter is significantly further away from the
integral-valent state. This tendency is a general result of the flow
equations approach and can be better understood in the context of a
characteristic energy scale (cf. Sec.~\ref{sec:Tk} below).  The
quasiparticle energies obtained from the 
numerical solution can be compared to those of the analytical
treatment in Sec.~\ref{sec:ana}. Their analytical expressions are given by
Eq.~\eqref{eq:ana1} and Eq.~\eqref{eq:anad}, which have proved to be formally
equivalent to the corresponding results of the SBMF theory. For the
expectation value $\langle P_i^0\rangle$ and the energy $\varepsilon_f^\ast$
we have used the numerically evaluated results in
Fig.~\ref{fig:dispersionEf9V3}. The graphs show a 
qualitatively good coincidence of both results for the excitation
energies. However, near 
the crossing point of $c$-like and $f$-like excitations the deviations become
most significant. Especially, the upper band
shows a slight depletion near the crossing point, which results in a
Van Hove-like singularity in the quasiparticle DOS. This feature is a result
of the consideration of a renormalisation of the energy $\varepsilon_f$ on a
large $\ell$-scale, which is driven by local fluctuations. The
analytical solution is based on the assumption of a fast flow of this energy
to its limit value, which is contrasted by Fig.~\ref{fig:ek3d} and
Fig.~\ref{fig:EfV_flow}. Here we perceive the characteristic $\ell$-scale of
all energies in the same region. However, the quasiparticle energies of the
analytical solution show a good correspondency to their numerical
counterparts.\par
Whereas the system in Fig.~\ref{fig:dispersionEf9V3} was classified as
mixed-valent system, Fig.~\ref{fig:dispersionEf9V2} is very close to
the integral-valent or Kondo behaviour. This manifests itself in the very
small expectation value $\langle P_i^0\rangle=0.083$ and the energy
$\varepsilon_f^\ast=1.003$ lying only slightly above the chemical
potential. This system also features an evident decrease of the gap
width. In comparison with that result the SBMF theory delivers $\langle
P_i^0\rangle_\mathrm{SB}=0.479$ and $\varepsilon_{f,\mathrm{SB}}^\ast=1.036$
indicating a 
mixed-valent behaviour. This distinctive tendency of the flow equation result
towards integral-valence is discussed  in connection with the Kondo
temperature in Sec.~\ref{sec:Tk} below.\par
\begin{figure}[h]
  \includegraphics[width=\linewidth]{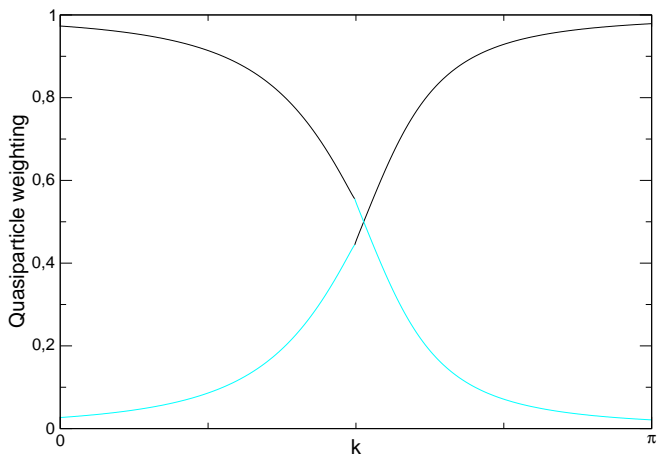}
  \caption{\label{fig:nk}Weighting factors $(\alpha_\mathbf{k}^\ast)^2$ (black
  solid line) and $\langle P_i^0\rangle (\beta_\mathbf{k}^\ast)^2$ for the
  same system as in Fig. \ref{fig:dispersionEf9V3}.}
\end{figure}

The drastic change in the character of the quasiparticles from $c$ to
$f$ character is a key result of the numerical calculations, as this
contributions evolve from the bare conduction electron band and the
dispersion-less $f$ level, respectively, during the
flow. This fact also ensures the right result in the limit of a system
without hybridisation. Notwithstanding, the different types of
excitations form two continuous bands. In the context of the SBMF
theory the side conditions are usually chosen in the way that each
quasiparticle bands carries a single character. The switch of the
excitation character in each band is also reflected by a change in the
behaviour of the weighting factors $\alpha_\mathbf{k}^\ast$ and
$\beta_\mathbf{k}^\ast$. The dispersion of their squares is shown in
Fig.~\ref{fig:nk}.
As the evaluation of the expectation values in
Sec.~\ref{sec:observ} incorporates both the quasiparticle energies and the
weighting factors, these quantities are not affected by the switch
of the excitation character.
\subsection{Kondo temperature}\label{sec:Tk}
Historically, the Kondo temperature $T_\mathrm{K}$ was introduced as a low energy scale
associated with the screening of a magnetic impurity in a metallic host. The
microscopic models for the description of such a situation are the single
impurity Anderson model (SIAM) and the single impurity Kondo model
(KLM). Renormalisation group calculations \cite{Wilson1975} as well as
SBMF theory \cite{SlaveBoson,Fulde1993} show a
dependence of this energy scale on the Kondo coupling
$J=(V^\mathrm{i})^2/|\varepsilon_f^\mathrm{i}-\mu|$ of the form
\begin{equation}\label{eq:TK_SIAM}
k_\mathrm{B}T_\mathrm{K}\propto \exp\biggl[-\frac{1}{J\nu_f\rho_0}\biggr]
\end{equation} 
This relation states, that this energy scale vanishes exponentially as
$V^\mathrm{i}$ goes to zero, a result that cannot be obtained by
ordinary perturbation
theory in $V^\mathrm{i}$. In a periodic array of magnetic impurities beside
the conventional screening of the local moments a coherence scale $T^\ast$
seem to exist, which is referred to in the literature as coherence temperature,
indicating the onset of Fermi-liquid behaviour. Whereas the Gutzwiller
calculations for the PAM suggest a significant enhancement of $T^\ast$
compared to the Kondo temperature of the SIAM, the SBMF theory provides a
coincidence of both temperatures for small $T^\ast$. The comparison of both
energy scales has been dedicated a controversial discussion. While some
numerical works show an enhancement of $T^\ast$ for systems at half filling
\cite{Pruschke2000}, a more qualitative discussion by Nozi\a'eres
\cite{Nozieres} relates this
issue to the so-called exhaustion problem. As Nozi\a'eres points out, only a
fraction of $n_\mathrm{scr}=\rho_0 T_\mathrm{K}$ conduction electrons 
contributes to the screening of local moments. As a consequence of this, he
deduces an upper bound of $T^\ast=\rho_0 T_\mathrm{K}^2/n_f$, where $n_f$ is
the number of magnetic impurities, for the lattice energy scale. This
behaviour is supported by numerical calculations by Vidhyadhiraja
\textit{et al.} \cite{Vidhyadhiraja2000}.\par
In accordance to the SBMF theory \cite{Fulde1993} we define the Kondo
temperature by the difference of the renormalised $f$ level and the chemical
potential as 
\begin{equation}\label{eq:Kondo_temperature}
  k_\mathrm{B}T^\ast=\varepsilon_f^\ast-\mu
\end{equation}
As we have mentioned at the beginning of this section, our method provides
valid results in a certain parameter regime. By lowering the hybridisation
$V^\mathrm{i}$ or by increasing the difference $\mu-\varepsilon_f^\mathrm{i}$
the self-consistency scheme breaks down as the expectation value $\langle
P_i^0\rangle$ vanishes. The vanishing point coincides with
$\varepsilon_f^\mathrm{i}$ slipping below the chemical potential, which is
consistent with the assumption of positive results for the lattice Kondo
temperature. \par
The evaluation of the Kondo temperature also allows us to estimate the
resulting effective masses of the quasiparticles by comparing this energy
scale to the band width of the bare conduction electrons. This leaves us with
the relation $m^\ast/m\sim W/k_\mathrm{B}T^\ast$. If we consider the system in
Fig.~\ref{fig:dispersionEf9V3}, we obtain an effective quasiparticle mass of
$m^\ast/m\sim 60$, indicating the system being in the mixed-valence regime. The
enhancement of the quasiparticle mass is due to the flatness of the
quasiparticle band near the Fermi surface, and it is also associated with a
large quasiparticle DOS. For the integral-valence system in
Fig.~\ref{fig:dispersionEf9V2} we even deduce an effective mass of $m^\ast/m\sim
600$. \par
\begin{figure}[h]
  \includegraphics[width=\linewidth,clip=]{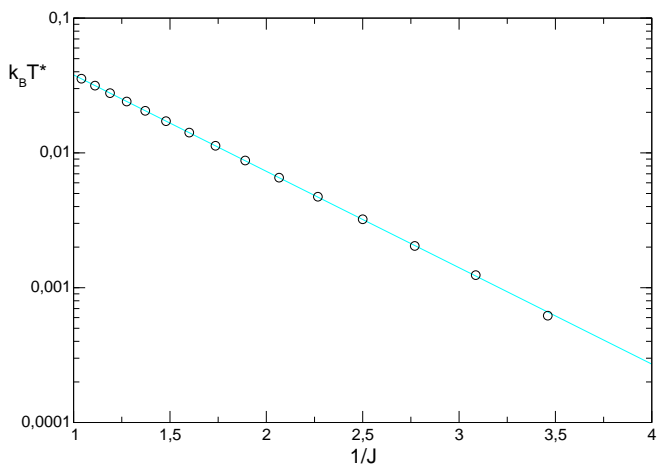}
  \caption{\label{fig:TkJ} Kondo Temperature as a function of $1/J$ (see text)
  for a system with $\varepsilon_f^\mathrm{i}\!=\!0.9$ and $\nu_f\!=\!2$. The
  blue solid line represents the fit $\beta \exp(-\delta/J\nu_f\rho_0)$ with
  $\beta\!=\!0.20$ and $\delta\!=\!1.65$.}
\end{figure}

To address the issue of the functional trend of the Kondo temperature for the
numerical solution, we evaluate this quantity by varying the coupling $J$. In
particular, we examine the dependence of $T^\ast$ on the hybridisation
$V^\mathrm{i}$. The result for the system  with $\varepsilon_f^\mathrm{i}=0.9$
is shown in Fig. \ref{fig:TkJ}. 
Here $V^\mathrm{i}$ is gradually decreased
until the solution breaks down below the value $V^\mathrm{i}=0.11$. The
numerical results are represented by circles, and the blue solid line
characterises the fit $\beta \exp(-\delta/J\nu_f\rho_0)$. Our findings for the
fit parameters are $\beta=0.20$ and $\delta=1.65$. 
The latter identifies the
value of the exponent and is significantly enhanced to the corresponding value
for the SIAM, i.~e. $\delta=1$. That means that the lattice energy scale features a
remarkable reduction. As we have mentioned above, this behaviour gives credence
to the exhaustion scenario.
\subsection{Green's functions}\label{sec:Green}
\begin{figure}[h]
  \includegraphics[width=\linewidth,clip=]{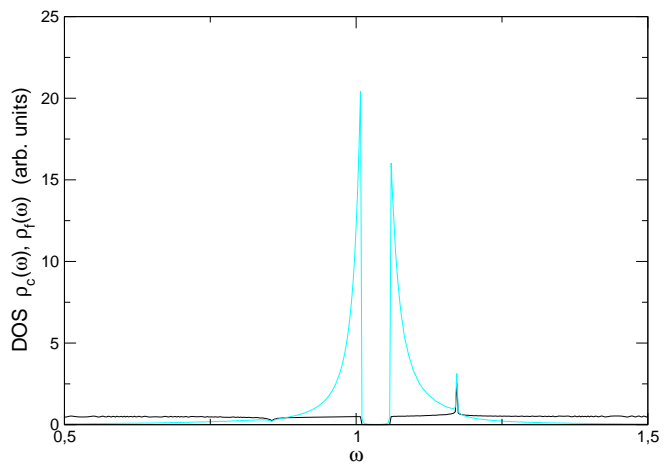}
  \caption{\label{fig:DOSEf9V3} Densities of state $\rho_c(\omega)$ (black
  solid line) and $\rho_f(\omega)$ (blue solid line) for the same system as in
  Fig. \ref{fig:dispersionEf9V3}.}
\end{figure}

In Sec.~\ref{sec:FE} we have explained the evaluation 
of correlation functions within the framework of the flow equation method. As such 
quantities are invariant with respect to the unitary transformation,
we are in the advantageous position of 
describing them with respect to a simple effective Hamiltonian. In this
context we use the results for the transformed operators from
Sec.~\ref{sec:observ}. For the conduction electrons this yields
\begin{align}\label{eq:Greenc}
  \langle \langle
  c_{\mathbf{k}\sigma}^{\phantom{\dagger}};
c_{\mathbf{k}\sigma}^\dagger\rangle\rangle(\omega+i\delta)  
  =&\frac{(\alpha_\mathbf{k}^\ast)^2}{\omega+i\delta -
  \varepsilon_\mathbf{k}^\ast}\nonumber\\
  &+\frac{\langle P_i^0\rangle (\beta_\mathbf{k}^\ast)^2}{\omega+i\delta -
  (\varepsilon_f^\ast + \langle P_i^0\rangle \Delta^\ast_\mathbf{k})}
\end{align}
Likewise we arrive at a representation for the $f$-electron
Green's function:
\begin{align}
  \langle\langle \hat{f}_{\mathbf{k}\sigma}^{\phantom{\dagger}};
  \hat{f}_{\mathbf{k}\sigma}^\dagger\rangle\rangle(\omega+i\delta)=&
 \langle P_i^0\rangle\frac{1- (\alpha_\mathbf{k}^\ast)^2}{\omega -
  \varepsilon_\mathbf{k}^\ast}\nonumber\\
  &+\langle P_i^0\rangle \frac{(\alpha_\mathbf{k}^\ast)^2}{\omega -
  (\varepsilon_f^\ast + \langle P_i^0\rangle \Delta^\ast_\mathbf{k})} 
\end{align}
The electronic structure can be made descriptive by evaluating the DOS of the
$c$ and $f$ electrons. These quantities are given by $\rho_c(\omega)=-1/(\pi
N)\sum_\mathbf{k} \mathrm{Im}\langle\langle
c_{\mathbf{k}\sigma}^{\phantom{\dagger}};
c_{\mathbf{k}\sigma}^\dagger\rangle\rangle(\omega+ i\delta)$ and similarly for
the $f$ DOS.
The results are shown in Fig. \ref{fig:DOSEf9V3} for the same
system as in Fig. \ref{fig:dispersionEf9V3}.
Here we have used Lorentzian
broadening with a width of $s=10^{-4}$. The DOS of the $c$ electrons remains
mainly flat as we expect from the constant DOS for the bare conduction
electrons, but it shows a gap around the renormalised energy
$\varepsilon_f^\ast$. On the other hand, the DOS of the $f$ electrons increases
sharply around the gap, meaning that there is a high DOS at the Fermi energy
($\mu=1$). This picture qualitatively coincides with the electronic structure
obtained from other approaches to the PAM, such as the SBMF theory, and
conveys the idea of heavy quasiparticles. A
remarkable difference to the SBMF result is an additional excitation
peak in the upper band, which 
is due to the above discussed deformation of the quasiparticle bands
(cf.~Fig.~\ref{fig:dispersionEf9V3}). It possesses the character of a Van
Hove-like singularity, as the DOS becomes very small in this energy
range. This large DOS comes from the consideration of local fluctuations,
which lead to a renormalisation of the $f$ level. This can be contrasted to
the case, in which we keep this energy constant during the flow process. With
this assumption the quasiparticle bands coincide with those of the SBMF
theory and do not show this type of excitation.
\section{Conclusions}
In this paper we have presented the description of heavy fermion behaviour
within the framework of Wegner's flow equation method. This work provides a
new semi-analytical approach to HFS. The basic idea of this scheme is the
derivation of an effective Hamiltonian, that describes the essential physics
of the system. Though this Hamiltonian is not diagonal in terms of usual
fermion creation and destruction operators, it characterises two decoupled
subsystems of electrons. Due to the Hubbard operators we have used for the $f$
electrons, the according subsystem needs further approximations for the
evaluation of physical properties, such as expectation values and correlation
functions. In general, our findings confirm the picture that is provided by
previous calculations, such as the SBMF theory. However, the flow equation
approach to the PAM yields an additional excitation in the upper quasiparticle
band and a significant decrease of the Kondo temperature.\par
In order to obtain a closed set of differential equations, we had to resort to
a decoupling approximation, and we have only considered contributions in the
leading order of $\nu_f$. The idea of the second step is similar to the
procedure used in the SBMF theory. However, in this context the two approaches
cannot be compared in a one-to-one fashion. As we have seen in the
discussion of Sec.~\ref{sec:one-particle} an improvement of the present
approximations can be obtained by an enhancement of the factorisation scheme
due to the importance of charge fluctuations in the case of small
degeneracy. An important difference of the
flow equation approach to the SBMF theory is the determination of the energy
$\varepsilon_f^\ast$. Whereas in the SBMF theory this energy is obtained by a
mean field procedure and has to be determined self-consistently, the present
method uses the integration of a differential equation. The procedure of
self-consistency within the flow equation method only refers to expectation
values. As we have shown in this paper, the differential equations can be
solved both analytically and numerically. Though the first was performed on
the basis of further simplifications, and it was intended to give a rough
estimate, it turned out to coincide qualitatively good with the numerical
solution. In accordance to the SBMF theory the present
calculations show a restriction to parameter regimes where the
occupation number of the $f$-sites is less than unity, as the
solution breaks down in the limit $\langle
P_i^0\rangle =0$. However, we obtain good results for quite small
deviations from this point. As we have stressed in
Sec.~\ref{sec:one-particle} this breakdown is a result of the
scheme of selfconsistency, while the integration of the flow
equations is rather robust and shows a good convergence even in this
limit.\par  
In the present paper we studied the paramagnetic phase of the
one-dimensional system. From recent works we know that the ground
state of the KLM in one dimension is either ferromagnetic or
non-ordered. On the other hand DMRG studies for the $U=\infty$ PAM also
showed regions of short-range antiferromagnetism in the Kondo
regime~\cite{Guerrero}, whereas the ground state is found to be
paramagnetic for the mixed-valent region. For the latter our results
are consistent to those obtained in Ref.~\onlinecite{Guerrero}. Near the
Kondo regime magnetic interactions, such as the RKKY interaction,
become more important. For a better understanding these
contributions must be incorporated in the present scheme. Due to the 
semi-analytical character of the flow equation method we believe
that this task could be accomplished for larger system sizes as
usually considered in the DMRG. Another intriguing issue which was
addressed in recent works about one-dimesional heavy fermion systems
is the study of Luttinger liquid
behavior~\cite{Shibata1997}. However, this subject was   
merely discussed for the KLM and an occurence of a Luttinger liquid in
the PAM is still an outstanding problem. The examination of this issue
though needs a further enhancement of the calculations and cannot be
obtained on a simple mean-field level.
\par
Our findings for the excitation of the quasiparticles allow a clear
distinction of $c$ and $f$ contributions. Generally, the electronic structure
of the PAM comprises two quasiparticle 
bands, which are divided by a  gap. The DOS is strongly enhanced at the Fermi
level, resulting in quasiparticle masses up to several hundreds of bar
electron masses. Such a behaviour characterises the experiments for heavy
fermion compounds. In comparison to previous theoretical works on the PAM, the
present paper shows an additional Van Hove-like excitation peak in the upper
quasiparticle band, which is due to a significant flattening of the
dispersion. \par
We also have investigated the dependence of the Kondo temperature on the
hybridisation $V^\mathrm{i}$. This quantity vanishes exponentially for small
values of $V^\mathrm{i}$. Our findings for $T^\ast$ are consistent with the
qualitative arguments proposed by Nozi\a'eres, suggesting a significant
decrease of $T^\ast$ compared to the Kondo temperature of the SIAM. 
\begin{acknowledgments}
The author gratefully acknowledges K.~Becker, R.~Hetzel, T.~Sommer and
M.~Vojta for fruitful discussions and helpful hints.
\end{acknowledgments}
\appendix
\section{Decoupling scheme}\label{sec:decoupling}
Here we present the factorisation procedure discussed in section \ref{sec:flow}
for the contributions in Eq. \eqref{eq:commutator}. With the
denotations $\langle c_{\mathbf{k}\sigma}^\dagger
c_{\mathbf{k}\sigma}^{\phantom{\dagger}}\rangle= \langle
n_{\mathbf{k}\sigma}^c \rangle$ and $\langle
\hat{f}_{i\sigma'}^\dagger\hat{f}_{i\sigma}^{\phantom{\dagger}}\rangle=\langle
n_{i\sigma}^f\rangle$ we obtain 
\begin{align}\label{eq:factor1}
\frac{1}{N}\sum_{\mathbf{kk'}}&\sum_{i\sigma}
e^{i(\mathbf{k}'-\mathbf{k})\mathbf{R}_i}\,
\eta_\mathbf{k} V_{\mathbf{k}'}
\,c_{\mathbf{k}\sigma}^\dagger
c_{\mathbf{k}'\sigma}^{\phantom{\dagger}}
P_i^0+\text{h.~c.}\nonumber\\
\longrightarrow&\;
2 \sum_{\mathbf{k}\sigma} \eta_\mathbf{k}
V_\mathbf{k}\langle P_i^0\rangle \,c_{\mathbf{k}\sigma}^\dagger
c_{\mathbf{k}\sigma}^{\phantom{\dagger}} +
\frac{2}{N} \sum_{i\mathbf{k}\sigma} \eta_\mathbf{k}V_\mathbf{k} \langle
n_{\mathbf{k}\sigma}^c\rangle\,P_i^0
\nonumber\\
& -2 \sum_{\mathbf{k}\sigma} \eta_\mathbf{k}
V_\mathbf{k}\langle P_i^0\rangle \,\langle n_{\mathbf{k}\sigma}^c \rangle
\end{align}
\begin{align}
\frac{1}{N}&\sum_{\mathbf{kk'}}\sum_{i\sigma\sigma'}
e^{i(\mathbf{k}'-\mathbf{k})\mathbf{R}_i}\,
\eta_\mathbf{k} V_{\mathbf{k}'}\, 
c_{\mathbf{k}\sigma}^\dagger 
c_{\mathbf{k}'\sigma'}^{\phantom{\dagger}}
\hat{f}_{i\sigma'}^\dagger\hat{f}_{i\sigma}^{\phantom{\dagger}}
+\text{h.~c.}\nonumber\\
\longrightarrow&\;
2 \sum_{\mathbf{k}\sigma} \eta_\mathbf{k}
V_\mathbf{k}\langle n_{i\sigma}^f\rangle \,c_{\mathbf{k}\sigma}^\dagger
c_{\mathbf{k}\sigma}^{\phantom{\dagger}} +
\frac{2}{N} \sum_{i\mathbf{k}\sigma} \eta_\mathbf{k}V_\mathbf{k} \langle
n_{\mathbf{k}\sigma}^c\rangle\,
\hat{f}_{i\sigma}^\dagger
\hat{f}_{i\sigma}^{\phantom{\dagger}}
\nonumber\\
&- 2 \sum_{\mathbf{k}\sigma} \eta_\mathbf{k}
V_\mathbf{k}\langle n_{i\sigma}^f\rangle \,\langle n_{\mathbf{k}\sigma}^c\rangle
\end{align}
In order to obtain the appropriate contributions to the one-particle
terms of the Hamiltonian from Eq. \ref{eq:factor1} it is convenient to
make use of the relation 
$P_i^0+\sum_\sigma
\hat{f}_{i\sigma}^\dagger\hat{f}_{i\sigma}^{\phantom{\dagger}}=\mathbf{1}$. The
remaining interactions are decoupled as follows
\begin{align}
\frac{1}{\sqrt{N}}\sum_{\mathbf{k}\sigma}\sum_{i\neq j}&
\eta_\mathbf{k} t_{ij} e^{-i\mathbf{kR}_i}
P_i^0
c_{\mathbf{k}\sigma}^\dagger\hat{f}_{j\sigma}^{\phantom{\dagger}}+\text{h.~c.}\nonumber\\
\longrightarrow&\;
\sum_{\mathbf{k}\sigma}\eta_\mathbf{k} \Delta_\mathbf{k} \langle
P_i^0\rangle \bigl(
c_{\mathbf{k}\sigma}^\dagger\hat{f}_{\mathbf{k}\sigma}^{\phantom{\dagger}}
+\hat{f}_{\mathbf{k}\sigma}^\dagger c_{\mathbf{k}\sigma}^{\phantom{\dagger}}
\bigr)\nonumber\\
&+\frac{1}{N}\sum_{i\mathbf{k}\sigma}\eta_\mathbf{k}\Delta_\mathbf{k}
\langle A_{\mathbf{k}\sigma}\rangle
P_i^0 \nonumber\\
&- \frac{1}{N}\sum_{i\mathbf{k}\sigma}\eta_\mathbf{k}\Delta_\mathbf{k}
\langle A_{\mathbf{k}\sigma}\rangle\,
\langle P_i^0\rangle
\end{align}
\begin{align}
\frac{1}{\sqrt{N}}
\sum_{\mathbf{k}\sigma\sigma'}\sum_{i\neq j}&
\eta_\mathbf{k}\,t_{ij}\,e^{-i\mathbf{kR}_i} 
\hat{f}_{i\sigma'}^\dagger\hat{f}_{i\sigma}^{\phantom{\dagger}}
c_{\mathbf{k}\sigma}^\dagger \hat{f}_{j\sigma'}^{\phantom{\dagger}} +
\text{h.~c.}\nonumber\\
\longrightarrow&\;
\sum_{\mathbf{k}\sigma}\eta_\mathbf{k} \Delta_\mathbf{k} \langle
n_{i\sigma}^f\rangle \bigl(
c_{\mathbf{k}\sigma}^\dagger\hat{f}_{\mathbf{k}\sigma}^{\phantom{\dagger}}
+\hat{f}_{\mathbf{k}\sigma}^\dagger c_{\mathbf{k}\sigma}^{\phantom{\dagger}}
\bigr)\nonumber\\
&+\frac{1}{N}\sum_{i\mathbf{k}\sigma}\eta_\mathbf{k}\Delta_\mathbf{k}
\langle A_{\mathbf{k}\sigma}\rangle
\hat{f}_{i\sigma}^\dagger
\hat{f}_{i\sigma}^{\phantom{\dagger}}\nonumber\\
&- \sum_{\mathbf{k}\sigma}\eta_\mathbf{k}\Delta_\mathbf{k}
\langle n_{i\sigma}^f\rangle
\langle A_{\mathbf{k}\sigma}\rangle
\end{align}
where $\langle A_{\mathbf{k}\sigma}\rangle=\langle
c_{\mathbf{k}\sigma}^\dagger\hat{f}_{\mathbf{k}\sigma}^{\phantom{\dagger}}
+ \hat{f}_{\mathbf{k}\sigma}^\dagger c_{\mathbf{k}\sigma}^{\phantom{\dagger}}
\rangle$ was used. 
\section{local expectation values}\label{sec:local_expectation}
The expectation value $\langle P_i^0\rangle$ is best evaluated by making use
of the the relation 
\begin{equation}
P_i^0=\mathbf{1} - \sum_\sigma \hat{f}_{i\sigma}^\dagger
\hat{f}_{i\sigma}^{\phantom{\dagger}} 
\end{equation}
Further we can take advantage of the translations invariance of the system,
which allows us to write $\langle
\hat{f}_{i\sigma}^\dagger \hat{f}_{i\sigma}^{\phantom{\dagger}}
\rangle\!=\!1/N \sum_j \langle\hat{f}_{j\sigma}^\dagger
\hat{f}_{j\sigma}^{\phantom{\dagger}}\rangle$. Consequently, it is convenient
to regard the unitary transformation of the operator
\begin{equation}
F\!=\!1/N\sum_{i\sigma}
\hat{f}_{i\sigma}^\dagger \hat{f}_{i\sigma}^{\phantom{\dagger}}
\end{equation}
Its $\ell$-dependence can be described in the usual way by $\mathrm{d}
F(\ell)/\mathrm{d}\ell\!=\![\eta(\ell),F(\ell)]$, where the generator is given
by \eqref{eq:generator}. Thus the following ansatz for the operator is expedient
\begin{align}
 F(\ell)= &\frac{1}{N}\sum_{\mathbf{k}\sigma}\bigl(A^{(1)}(\ell) +
 A_\mathbf{k}^{(2)}(\ell) \bigr) \hat{f}_{\mathbf{k}\sigma}^\dagger
\hat{f}_{\mathbf{k}\sigma}^{\phantom{\dagger}}\nonumber\\
&+ \frac{1}{N}\sum_{\mathbf{k}\sigma} B_\mathbf{k} (\ell)
 c_{\mathbf{k}\sigma}^\dagger 
 c_{\mathbf{k}\sigma}^{\phantom{\dagger}}\nonumber\\
 &+\frac{1}{N}\sum_{\mathbf{k}\sigma} G_\mathbf{k}(\ell) \bigl(
 c_{\mathbf{k}\sigma}^\dagger
 \hat{f}_{\mathbf{k}\sigma}^{\phantom{\dagger}} +
 \hat{f}_{\mathbf{k}\sigma}^\dagger
 c_{\mathbf{k}\sigma}^{\phantom{\dagger}} \bigr) + \frac{E(\ell)}{N}
\end{align}
In this connection the boundary condition 
$A^{(1)}(\ell\!=\!0)=1$ hold, whereas all the other coefficients are equal to
zero for $\ell\!=\!0$. Evaluating the commutator we arrive at
\begin{widetext}
\begin{align}
  [\eta,F]=&-\frac{1}{N^2}\sum_{\mathbf{k}\sigma}\eta_\mathbf{k} \bigl(
  -B_\mathbf{k} +  A^{(1)} + \langle P_i^0 \rangle A_\mathbf{k}^{(2)}
  \bigr)\;\bigl(c_{\mathbf{k}\sigma}^\dagger
  \hat{f}_{\mathbf{k}\sigma}^{\phantom{\dagger}} +
  \hat{f}_{\mathbf{k}\sigma}^\dagger
  c_{\mathbf{k}\sigma}^{\phantom{\dagger}} \bigr)
  + \frac{1}{N}\sum_{i\mathbf{k}}\sum_{\sigma\sigma'} 2
  \eta_\mathbf{k} G_\mathbf{k}\langle
  n_{\mathbf{k}\sigma}^c\rangle
 \hat{f}_{{i}\sigma}^\dagger\hat{f}_{i\sigma}^{\phantom{\dagger}}  \nonumber\\
  &+\frac{1}{N}\sum_{\mathbf{k}\sigma} 2 \eta_\mathbf{k}G_\mathbf{k}
  \langle P_i^0 \rangle  
  c_{\mathbf{k}\sigma}^\dagger 
  c_{\mathbf{k}\sigma}^{\phantom{\dagger}}
  -\frac{1}{N}\sum_{\mathbf{k}\sigma} 2 \eta_\mathbf{k}G_\mathbf{k}
  \hat{f}_{\mathbf{k}\sigma}^\dagger
  \hat{f}_{\mathbf{k}\sigma}^{\phantom{\dagger}}
  +\frac{1}{N^2}\sum_{\mathbf{k}i\sigma} 2 \eta_\mathbf{k} G_\mathbf{k}\langle
  n_{\mathbf{k}\sigma}^c\rangle \bigl(1-\langle P_i^0\rangle\bigr)
  \end{align}
\end{widetext}
Here the decoupling approximation, which was discussed in Sec. \ref{sec:flow}
is already done. As a result we obtain a closed set of flow equations for the coefficients
\begin{align}
\frac{\mathrm{d}A^{(1)}}{\mathrm{d}\ell}=&-\frac{1}{N}
\sum_{\mathbf{k}\sigma}2\langle n_{\mathbf{k}\sigma}^c\rangle
\eta_\mathbf{k}G_\mathbf{k}\\
\frac{\mathrm{d}A^{(2)}}{\mathrm{d}\ell}=&-2\eta_\mathbf{k}G_\mathbf{k}\\
\frac{\mathrm{d}B_\mathbf{k}}{\mathrm{d}\ell}=&2\langle P_i^0\rangle
\eta_\mathbf{k}G_\mathbf{k}\\ 
\frac{\mathrm{d}G_\mathbf{k}}{\mathrm{d}\ell}=&-\bigl(B_\mathbf{k}-A^{(1)}-\langle
P_i^0\rangle A_\mathbf{k}^{(2)}\bigr)\eta_\mathbf{k}\\
\frac{\mathrm{d}E}{\mathrm{d}\ell}=&\sum_{\mathbf{k}\sigma}2
\eta_\mathbf{k}G_\mathbf{k}\langle n_{\mathbf{k}\sigma}^c\rangle
\bigl(1-\langle P_i^0\rangle\bigr) 
\end{align}
The following exact relations can be derived
\begin{align}
  A^{(1)}=&1+\frac{1}{N}\sum_{\mathbf{k}\sigma} \langle
  n_{\mathbf{k}\sigma}^c\rangle A_\mathbf{k}^{(2)} \label{eq_A1}\\ 
  A_\mathbf{k}^{(2)}=&- \frac{B_\mathbf{k}}{\langle P_i^0\rangle}\\
  E=&N\bigl(1-\langle P_i^0\rangle \bigr)\bigl(1- A^{(1)}\bigr)
\end{align}
Furthermore the differential equation
\begin{equation}
  \langle P_i^0\rangle \frac{\mathrm{d}G_\mathbf{k}^2}{\mathrm{d}\ell} 
  + \frac{\mathrm{d}B_\mathbf{k}^2}{\mathrm{d}\ell} - A^{(1)}
  \frac{\mathrm{d}B_\mathbf{k}}{\mathrm{d}\ell} =0\label{DGL_Gk}
\end{equation}
holds. As a simplification we regard the summation on the right hand side of
\eqref{eq_A1} a small correction, and we neglect it. Such an approximation is
well confirmed by numerical simulations. As a consequence we obtain
$A^{(1)}\!=\!1$ for the entire flow, and the differential equation
\eqref{DGL_Gk} can be integrated straightforward
\begin{equation}
  \frac{\mathrm{d}}{\mathrm{d}\ell} \arcsin \bigl[
  1-2B_\mathbf{k}\bigr]=-
  2 \sqrt{\langle P_i^0\rangle} \eta_\mathbf{k}
\end{equation}
With the help of the unitarity relation \eqref{eq:unitarity} and
\eqref{DGL:alpha} we can identify $B_\mathbf{k}=\langle P_i^0\rangle
\beta_\mathbf{k}^2$, and the expectation value thus reads
\begin{align}
  \langle P_i^0 \rangle =& 
  1-{}^\ast \langle F^\ast \rangle^\ast \nonumber\\
  =& 1- \frac{1}{N}\sum_{\mathbf{k}\sigma}
  \biggl[\bigl(1-(\alpha_\mathbf{k}^\ast)^2\bigr) \;{}^\ast\langle  
    c_{\mathbf{k}\sigma}^\dagger
    c_{\mathbf{k}\sigma}^{\phantom{\dagger}}\rangle^\ast\nonumber\\
    &\hspace{5em}+\bigl(1-(\beta_\mathbf{k}^\ast)^2\bigr)\;
         {}^\ast\langle \hat{f}_{\mathbf{k}\sigma}^\dagger
         \hat{f}_{\mathbf{k}\sigma}^{\phantom{\dagger}}\rangle^\ast
         \biggr]
\end{align}
An equivalent procedure can be used in order to derive the expectation values
$\langle c_{\mathbf{k}\sigma}^\dagger
c_{\mathbf{k}\sigma}^{\phantom{\dagger}}\rangle$  and $\langle
A_{\mathbf{k}\sigma}\rangle$. Due to their non-local character the results
coincide with those obtained in Sec. \ref{sec:flow}. Therefore it is more 
reasonable to use the single operator transformation, because this approach
can be used to describe correlation functions, as pointed out in Sec. \ref{sec:Green}. 
%


\begin{thebibliography}{99}
\bibitem{Flowparameter} in case we apply Wegner's choise for the generator, namely
  $\eta=[H_0,H_1]$, the flow parameter $\ell$ has the dimension of an inverse
  energy squared
\bibitem{nuf} here we have again neglected contributions of order $\nu_f^{-1}$
\end{thebibliography}
\end{document}